\documentclass[preprint]{elsarticle}

\usepackage[utf8]{inputenc}
\usepackage{graphicx}
\graphicspath{{Plots/}}
\usepackage{amsmath}
\usepackage[margin=1in]{geometry}
\usepackage{float} 
\usepackage{wrapfig}
\usepackage{multirow}
\usepackage{siunitx}
\usepackage{url}
\usepackage{array}
\usepackage{subfig} 
\usepackage[moderate]{savetrees}
\newcolumntype{L}{>{\centering\arraybackslash}m{3cm}}
\usepackage{amssymb}
\usepackage{bm}
\usepackage{subfig}
\usepackage{makecell}
\usepackage{array}
\usepackage{caption}
\usepackage{multirow}
\newcolumntype{P}[1]{>{\centering\arraybackslash}p{#1}}
\usepackage{siunitx}
\usepackage{subfig}
\usepackage{gensymb}
\usepackage{miller}
\usepackage{lmodern}
\usepackage[T1]{fontenc}
\usepackage{bold-extra}

\usepackage[ruled,vlined]{algorithm2e} % For formatting algorithms

\SetKwInput{KwInput}{Input}                % Set the Input
\SetKwInput{KwOutput}{Output}

\SetCommentSty{mycommfont} 
\DontPrintSemicolon



\usepackage[table]{xcolor}
\usepackage{xcolor}

\usepackage{todonotes}

\usepackage[english]{babel}

\usepackage{fancyhdr}

\usepackage{lineno,hyperref}
\usepackage{multirow}
\usepackage{siunitx}
\usepackage{graphicx}
\usepackage{subfig}
\usepackage{makecell}
\usepackage{array}
\usepackage{caption}
\usepackage{float} 
\usepackage{amssymb}
\usepackage[margin=1in]{geometry}
\newcolumntype{P}[1]{>{\centering\arraybackslash}p{#1}}

\newcommand{\fref}[1]{\emph{fig}.~\ref{#1}}%
\newcommand{\Fref}[1]{Fig.~\ref{#1}}%
\newcommand{\eref}[1]{\emph{eq}.~(\ref{#1})}%
\def\esref[#1]#2{\emph{eq}.~(\ref{#2}#1)}%
\newcommand{\cref}[1]{\emph{chapter}~\ref{#1}}%
\newcommand{\Cref}[1]{Chapter~\ref{#1}}%
\newcommand{\aref}[1]{\emph{Algorithm}~\ref{#1}}%
\newcommand{\limsum}[2]{\displaystyle\sum\limits_{#1}^{#2}}
\newcommand{\algo}{Or\textsc{isodata}}

\newcommand{\ovito}{\textsc{Ovito}}
\newcommand{\isodata}{\textsc{isodata}}

\begin{document}
	
\begin{frontmatter}

\title{Grain segmentation in atomistic simulations using \\orientation-based iterative self-organizing data analysis}

\author[mimm]{M.~Vimal}
\author[mimm,fzj,rwth]{S.~Sandfeld}
\author[mimm]{A.~Prakash\corref{cor1}}
\ead{arun.prakash@imfd.tu-freiberg.de}
%	\ead[url]{}
		
\address[mimm]{Micromechanical Materials Modelling (MiMM), \\
Institute of Mechanics and Fluid Dynamcis, \\TU Bergakademie Freiberg, Lampadiusstra{\ss}e 4, 09599 Freiberg, Germany}
\address[fzj]{Institute for Advanced Simulation -- IAS-9: Materials Data Science and Informatics, \\Forschungszentrum Juelich GmbH, 52425 Juelich, Germany}
\address[rwth]{Chair of Materials Data Science and Materials Informatics,\\ Faculty 5 -- Georesources and Materials Engineering, \\RWTH Aachen University, 52056 Aachen, Germany}

\cortext[cor1]{Corresponding author}

\begin{abstract}

Atomistic simulations have now established themselves as an indispensable tool in understanding deformation mechanisms of materials at the atomic scale. Large scale simulations are regularly used to study the behavior of polycrystalline materials at the nanoscale. In this work, we propose a method for grain segmentation of an atomistic configuration using an unsupervised machine learning algorithm that clusters atoms into individual grains based on their orientation. The proposed method, called the \algo{} algorithm, is based on the iterative self-organizing data analysis technique and is modified to work in the orientation space. The working of the algorithm is demonstrated on a 122 grain nanocrystalline thin film sample in both undeformed and deformed states. The \algo{} algorithm is also compared with two other grain segmentation algorithms available in the open-source visualization tool \ovito{}. The results show that the \algo{} algorithm is able to correctly identify deformation twins as well as regions separated by low angle grain boundaries. The model parameters have intuitive physical meaning and relate to similar thresholds used in experiments, which not only helps obtain optimal values but also facilitates easy interpretation and validation of results.

\end{abstract}

\begin{keyword}
\texttt{Atomistic simulations, Unsupervised machine learning, orientation-based \textsc{isodata} clustering, grain segmentation, graph clustering, minimum spanning tree}

\end{keyword}

\end{frontmatter}

%\linenumbers

\newpage
\section{Introduction}
Polycrystalline materials, made of aggregates of single crystals or grains, constitute a large fraction of materials used today. The properties of such materials depend on the constituent microstructure, comprising of topological entities such as grain interior, grain boundaries (GBs), triple junctions and quadruple points. A well-known example is the Hall-Petch equation which relates the yield stress in the material to its mean grain size \cite{hall1951}.

Of particular interest have been nanocrystalline (NC) materials, with mean grain sizes in the range of $10\sim 100$ nm. The increased interest in NC materials in recent years has been spurred both by advances in processing and by insights obtained via computations \cite{kumar2003,wolf2005,meyers2006,vanswygenhoven2006}. Atomistic simulations of the molecular dynamics/statics kind have played a key role in elucidating deformation mechanisms in NC materials \cite{farkas2013,hahn2015}. Such simulations have now clearly detailed the role of GBs in NC materials. For instance, the negative Hall-Petch effect observed in almost all NC materials at very small grain sizes is due to a transition of deformation mechanism from dislocation mediated plasticity to that dominated by grain boundary sliding \cite{schiotz1998,vanswygenhoven2001}. Furthermore, GBs not only act as nucleation sites for dislocations due to the absence of intra-granular sources like Frank-Read or spiral sources, defect structures at GBs such as ledges act as effective pinning points for nucleated and/or propagating dislocations \cite{schiotz2004,vanswygenhoven2006b,
panzarino2016plasticity,prakash2017influence}.

It hence follows that understanding the influence of GBs and related topological entities is the key to obtaining structure property relationships for polycrystalline materials. An important task to this end in atomistic simulations is the partitioning of atoms into grain interior and grain boundary regions, and furthermore, to identify and track individual grains. Conventional characterization methods, like common neighbor analysis (CNA) \cite{honeycutt1987molecular, faken1994systematic}, centrosymmetry parameter (CSP)  \cite{kelchner1998dislocation}, coordination analysis (CA) based on the local structural environment around an atom, or filtering methods based, e.g., on the potential energy of the atom, can be used to identify local crystal structure and assign atoms a corresponding label as ``\emph{bulk}'' or ``\emph{defect}'' atoms. Such an approach is taken by Tucker and Foiles \cite{tucker2013} in their algorithm where fcc atoms (determined by CNA) are designated as \emph{grain-center} or \emph{grain-edge} atoms. But they are not universally applicable: CNA, CSP are not suitable for cases with low angle GBs; CA does not identify coherent twin boundaries.

Identification of grains hence requires calculation of and accounting for an inherent property: orientation. The grain tracking algorithm (GTA) developed by Panzarino \emph{et al.}\cite{panzarino2014tracking,panzarino2015quantitative}, involves the calculation of a local per atom crystallographic orientation using the geometry of the unit cell (obtained by CNA and CSP). This is followed by an iterative process to identify individual grains as contiguous regions where the misorientation between nearest neighbors of atoms is less than a pre-defined threshold. A similar approach is taken by Hoffrogge \emph{et al.} \cite{hoffrogge2017grain} who suggest an additional ``\emph{global}'' criterion to track the misorientation to the mean orientation of the grain. Such methods are, however, sensitive to local perturbations in the structural environment resulting from, e.g., strain or temperature.

Partitioning of a dataset into different clusters, e.g. grains, is a well researched unsupervised machine learning problem. The idea is to form a cluster with data points that are as similar as possible to each other within the cluster, whilst being as different as possible to data points in a different cluster. The open-source visualization tool \ovito{} \cite{ovito} provides implementations of two hierarchical clustering methods in its grain segmentation modifier. The first approach is similar to the GTA of Panzarino \emph{et al.}\cite{panzarino2014tracking,panzarino2015quantitative}, but uses the minimum spanning tree representation of the input structure, and additionally, computes the local atomic orientation using the polyhedral template matching \cite{larsen2016robust} algorithm. The second approach also uses graph clustering, but with different weights for the graph edges in comparison to the former. Grains are subsequently built by contracting graph edges using the node sampling method \cite{bonald2018hierarchical}. With its automatic mode, the algorithm chooses a good threshold value using a sequence of graph merging steps. The drawback, however, is that the threshold value (in both automatic and manual) has no intuitive physical meaning.

In this work, we propose an alternative approach that uses a centroid based partitioning technique and shows advantages to the graph clustering algorithms particularly for large datasets \cite{popat2014review, xu2015comprehensive}. Our approach, called \algo{} algorithm, is based on the iterative self organizing data analysis (\isodata{}) method \cite{ball1965isodata}, an unsupervised learning algorithm that is widely employed in remote sensing applications \cite{el2015hyperspectral,Jain88, manakos2000comparison,	abbas2016k}. An extension of the K-means algorithm, \isodata{} method has the advantage of automatically selecting the final number of clusters based on certain heuristics. The self organizing capabilities and the ability to \emph{split} clusters with larger spread and \emph{merge} similar clusters based on the thresholds are its key advantages.

The \algo{} algorithm retains the basic structure of the \isodata{} algorithm suggested by Ball \emph{et al.} \cite{ball1965isodata}, but modifies the \emph{split} and \emph{merge} procedures to account for the non-Euclidean nature of the orientation space. The method works for both undeformed and deformed states, and is able to identify both low angle GBs and twinned regions well. In any clustering problem, the threshold parameters can have a significant influence on the final clustering results and must be carefully chosen \cite{berkhin2006survey,rodriguez2019clustering}. Herein lies the advantage of the \algo{} algorithm: the intuitive nature of the threshold parameters, which essentially split and merge clusters using orientation spread and misorientation between clusters, help in obtaining optimal parameter values easily and ensure robust and reliable results. 

%The plan of this paper is as follows: in \sref{sec:orisodata_details}, we present the structure of our \algo{} algorithm in detail. The two grain segmentation algorithms implemented in \ovito{} are presented briefly in \sref{sec:ovito_algo_methods}. Results of testing and validation of \algo{} on a bicrystal sample is presented in \sref{sec:test_bicrystal}. In \sref{sec:test_polycrystal}, we present the results of \algo{} algorithm on a large thin film polycrystalline sample in both undeformed and deformed states, and compare the results with those from the grain segmentation algorithms in \ovito{}. A detailed discussion is subsequently presented in \sref{sec:discussion}.

\newpage
\section{Details of the algorithm}
\label{sec:orisodata_details}

The generic \isodata{} algorithm involves an iterative approach of splitting clusters based on the standard deviation of the data points along each dimension, and merging two clusters based on their Euclidean inter-cluster distance. A detailed description of the \isodata{} algorithm can be found in Ball \emph{et al.} \cite{ball1965isodata}. The overall idea is to reduce the variation in each cluster and to combine similar clusters over the iterations.

The basic steps involved in a generic \isodata{} clustering algorithm are,

\begin{enumerate}
	\item Randomly sample the initial centroids for the given initial number of clusters
	\item \label{item:sortStep} Sort data points based on their proximity to the cluster centroids 
	\item Recompute the cluster centroids
	\item Split clusters if the \emph{standard deviation} along any dimension is greater 	than the user defined split threshold
	\item Merge pairs of clusters if their \emph{inter-cluster Euclidean distance} is lower than the user-defined merge threshold 
	\item Go to step \ref{item:sortStep} until convergence is achieved
\end{enumerate}

\subsection{Orientation-based ISODATA clustering}
The use of standard deviation and the Euclidean distance metric in the generic ISODATA algorithm, as measures for variation within a cluster and similarity between clusters, respectively, is insufficient for identifying grains based on orientations, since orientations do not reside in Euclidean space. We hence propose a modified ISODATA algorithm ---the \algo{} algorithm--- by incorporating orientation-based metrics into the cluster split and merge procedures. The general structure of the main program that executes the \algo{} algorithm is shown in \aref{algo:main_program}.

\newpage
\begin{algorithm}[H]
	\SetAlgoLined
	\KwInput{
		\begin{enumerate}
			\item Atomistic configuration (AtomId, Position, Orientation)
			\item Lattice constant: $a_0$
			\item Lattice structure type: \emph{e.g. fcc/bcc/hcp}
			\item Per-atom volume: $V_i$
			\item Cluster split threshold angle (Maximum disorientation spread within a
			grain): $\Psi_{\textrm{split}}$
			\item Cluster merge threshold angle (Minimum disorientation angle between two
			grains): $\Phi_{\textrm{merge}}$
			\item Initial number of clusters: $n^{init}_{clus}$
			\item Minimum number of atoms per cluster \emph{or} Minimum grain size: $D^{min}_g$
			\item Maximum number of iterations: $n_{iter}$
			\item Convergence tolerance: \textsc{tol}
		\end{enumerate}
	}              % Set the Input
	\KwOutput{Clusters (Grains)}

	Data import\;
	Extract bulk atoms (using \textcolor{red}{Lattice structure type}) from the
	rest of the sample\;
	\underline{Or{\sc isodata} clustering} $\rightarrow$ Algorithm \ref{algo:isodata}
	\tcp*{Only the bulk atoms data are used as input for clustering}
	
	\For{$i\gets1$ \KwTo Number of orphan atoms}{
		Assign orphan atom $\rightarrow$ Nearest large cluster
	} 
		
	\underline{Merge clusters}  $\rightarrow$ Algorithm \ref{algo:merge_cluster} \label{algo:merge_cluster_pos}
	\tcp*{Iterative merge until no clusters get merged}
	
	\For{$j\gets1$ \KwTo No of clusters}{
		Calculate geometric centroid of each cluster
	} 
	
	Data export
	
	\caption{Main program}
	\label{algo:main_program}
\end{algorithm}

The input to the \algo{} clustering algorithm is a snapshot of an atomistic simulation, which contains the positions, local crystal structure and orientation of individual atoms.  We use the polyhedral template matching algorithm (PTM) \cite{larsen2016robust} as implemented in \textsc{Ovito} \cite{ovito} to determine the local orientation and atomic structure of each atom. Atomic volume is computed using a Voronoi tessellation. The structure of the \algo{} algorithm is shown in \aref{algo:isodata}.

\newpage

\begin{algorithm}[H]
	\SetAlgoLined
	\KwInput{ 1. Position and orientation of atoms, 2. Values of clustering parameters
	} 
	% Set the Input
	\KwOutput{1. Clusters, 2. Mean orientations of large clusters}

	Assign atoms to clusters randomly \tcp*{Number of clusters, $n_{clus}$ $=$ \textcolor{red}{$n^{init}_{clus}$}}
	
	Calculate mean orientation of each cluster \tcp*{L2-chordal mean and atomic volume as weights at every calculation}

	\For{$i\gets1$ \KwTo \textcolor{red}{$n_{iter}$}}{
		
		Calculate orientation distance of each atom to mean orientation of each individual cluster\;
		Assign atoms to the nearest cluster based on orientation distance\;
		Update mean orientation of each cluster\; %\tcp*{L2-Chordal mean and per atom	voronoi volume $\rightarrow$ weights}
		Ignore clusters with number of atoms $<$ \textcolor{red}{$D^{min}_g$} \tcp*{Data points  in the ignored clusters are not deleted
			but only ignored for the remaining iterations} \label{line:ignore_clus}
		
		Update array of mean orientations and $n_{clus}$ \tcp*{If number of ignored clusters $\geq$ 1}

		\textbf{Split clusters}  $\rightarrow$ Algorithm \ref{algo:split_cluster}
		%\tcp*{Misorientation based}
		
		\If{Clusters split}
		{	
			Update array of mean orientations and $n_{clus}$ \;
			Calculate orientation distance of each atom to the mean orientation of all clusters \;
			Assign atoms to the nearest cluster based on orientation distance\;
			Update array of mean orientations and $n_{clus}$
			
		}
		
		\textbf{Merge clusters}  $\rightarrow$ Algorithm \ref{algo:merge_cluster}
		%\tcp*{Misorientation based}
		
		Update array of mean orientations and $n_{clus}$ \tcp*{If any cluster gets merged}
		
		$\Delta$ mean orientation = Current mean orientation - Last mean orientation\;
		Last mean orientation $\leftarrow$ Current mean orientation \; 
		
		\If{$\bigg($i == \textcolor{red}{$n_{iter}$}$\bigg)$ \textbf{OR} $\bigg($absolute max ($\Delta$ mean orientation) $<=$ \textcolor{red}{\textsc{tol}}$\bigg)$}
		{	
			
			\For{$j\gets1$ \KwTo $n_{clus}$}{
				Position based cluster split \tcp*{mean of 1$^{st}$ and 2$^{nd}$ neighbor distance $\rightarrow$ Cut-off distance}%$ Positional
%					connectivity check by constructing an undirected graph with neighbors of the
%					atoms in each cluster} \label{line:pos_con}
				
			}
			
			\textbf{Split clusters}  $\rightarrow$ Algorithm \ref{algo:split_cluster}
			\tcp*{Iterative split until all the newly split (position based) clusters
				maintain their grain orientation spread (GOS) lower than $\Psi_{\textrm{split}}$}
			
			Append all the ignored clusters (line \ref{line:ignore_clus} ) to the array of current
			clusters

			%%%%%%%% Added from Algo.1 %%%%%%%%
			Designate clusters as large and orphan clusters based on \textcolor{red}{$D^{min}_g$}\;
			
			Calculate mean orientation of each large cluster %\tcp*{L2-Chordal mean and per atom voronoi volume $\rightarrow$ weights} \label{line:mean_orient}

			%%%%%%%% Added from Algo.1 %%%%%%%%
			
			\textbf{return} Clusters, mean orientations of clusters
			
		}  
		
%		\Else{
%			Last mean orientation $\leftarrow$ Current mean orientation
%		}
		
	} 
	
	\caption{Or\textsc{isodata} clustering}
	\label{algo:isodata}
\end{algorithm}
%\newpage

Briefly, the program works as follows: atoms in the snapshot are first classified into bulk and non-bulk atoms using the local atomic structure (here, fcc). Subsequently, the \algo{} algorithm (see \emph{Algorithm}~\ref{algo:isodata}) is applied on the bulk atoms, which organizes them into individual clusters. Herein, all clusters that have fewer than a predefined number of atoms are designated as small/orphan clusters. In an iterative procedure, large (non-orphan) clusters are then split and merged using user-defined thresholds for the same. The details of the split and merge routines are presented below. Subsequently, all atoms in orphan clusters and non-bulk atoms that were previously neglected, are assigned to a large cluster to which they are in close positional proximity. Finally, the average orientation and center of mass of each cluster are calculated.

%%%%%%%%%%%%%%%%%%%%%%%%%%% REWRITTEN %%%%%%%%%%%%%%%%%%%%%%%%%%%%%%%%%

The orientation distance metric used in the current work is the misorientation angle defined via the geodesic distance metric \cite{hartley2013rotation,huynh2009metrics} and is given by the dot product of two orientations (here, quaternions) $q_A$ and $q_B$ as follows:
\begin{equation}
\theta \ = \ 2 cos^{-1}\left( q_A \cdot q_B\right).
\end{equation}
Further details on the calculation of the orientation distance are provided in the supplementary material.

For calculating the mean orientation of each cluster, the chordal L2 mean (projected/induced arithmetic mean) \cite{hartley2011l1,moakher2002means} based on the cost minimization method is used. The cost function is given by 
\begin{equation}
C(R) \ = \limsum{i=1}{n} (R_i - R)^2\, ,
\end{equation}
where $R$ is the average orientation of a set of orientations ($R_i$) that minimizes the cost function $C(R)$.

After convergence, the final clusters are split based on their positional connectivity by analyzing the connected components in a sparse graph, constructed using the atomic positions of each cluster \cite{pearce2005improved} with the neighbors of each atom within a given cutoff distance. \Fref{fig:AlgoSchema}a shows the position based cluster split of 3 regions having identical orientations, as a result of which they would be classified to a single cluster. These regions are, however, not directly connected to each other. The average of the 1$^{st}$ and 2$^{nd}$ neighbor distance for the given lattice structure type is chosen as the neighbor cutoff distance (d\textsubscript{cutoff}) for evaluating the connectivity.

\subsection{Cluster split}

\begin{figure}
\centering
\includegraphics[width=0.7\textwidth]{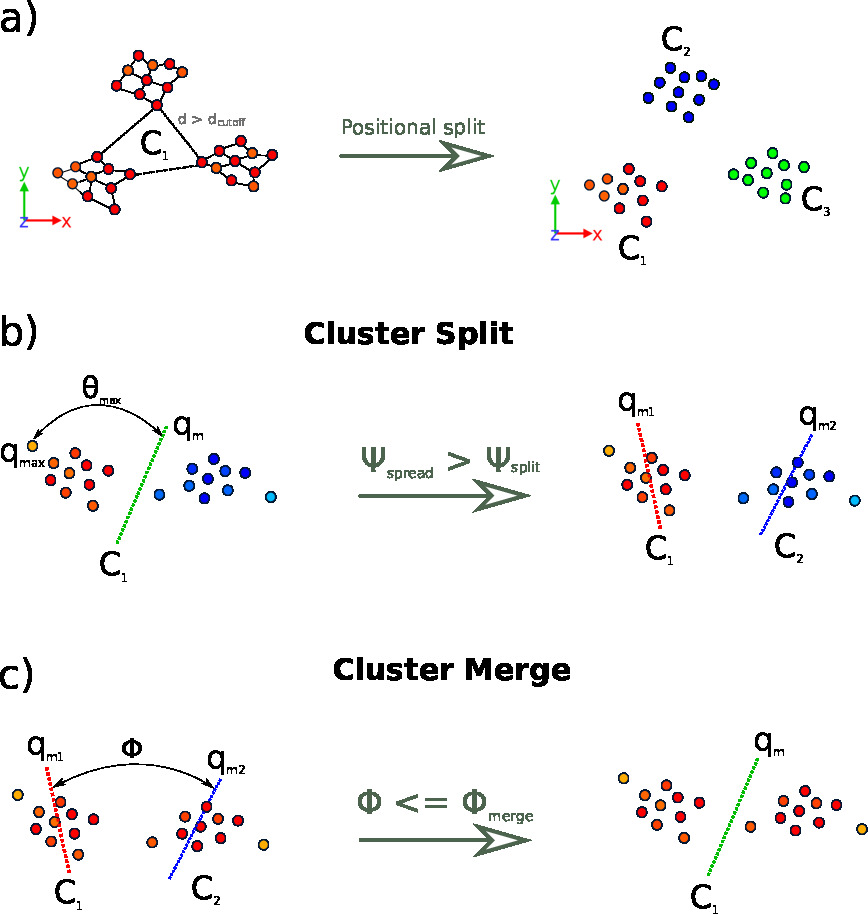}
\caption{Schematic representation of the a) positional split, b) cluster split and c) cluster merge processes.}\label{fig:AlgoSchema}
\end{figure}

The algorithm for splitting clusters using an orientation-based metric is shown in \aref{algo:split_cluster}. For computing the spread of orientations in each cluster, the grain orientation spread (GOS) is used. This metric is also used to characterize deformed microstructures using electron back-scatter diffraction (EBSD) techniques \cite{jorge2005study,mitsche2007recrystallization,ayad2012grain}. The GOS ($\Psi$\textsubscript{spread}) of a cluster is given by \cite{allain2012study}: 
\begin{equation}
\Psi\textsubscript{spread} \ = \ \frac{1}{N} \ \sum_{i} \ \theta_{q_i
	\leftrightarrow q_m},
\end{equation}
where $N$ is the number of orientations (atoms) in the cluster under consideration and ($\theta_{q_i \leftrightarrow q_m}$) is the orientation distance between the orientation ($q_i$) of the atom and mean orientation ($q_m$) of the cluster. 
\Fref{fig:AlgoSchema}b shows a schematic of a cluster with dissimilar orientations split into two different clusters.  All clusters satisfying the split condition, $\Psi$\textsubscript{spread}$>$ $\Psi$\textsubscript{split}, are split into two different clusters. 

\begin{algorithm}[H]
	\SetAlgoLined
	\KwInput{
		
		\begin{enumerate}
			
			\item Clusters 
			\item \textcolor{red}{$\Psi_{\textrm{split}}$}       
			% Set the Input
		\end{enumerate}
	}
	\KwOutput{Clusters after split}
	
	\SetKwProg{Fn}{Function}{ is}{end}
	\Fn{\textbf{get GOS}(Cluster)}{
		
		Calculate orientation distance of all atoms with respect to the mean orientation
		(q$_m$) of the given \textit{cluster}\;
		Find the corresponding orientation (q\textsubscript{max}) that has the maximum
		orientation distance ($\theta$\textsubscript{max})\;
		Calculate the GOS ($\Psi$\textsubscript{spread}) from
		the orientation distances
		
		\textbf{return} q\textsubscript{max}, $\Psi$\textsubscript{spread}\;}

	\For{$i\gets1$ \KwTo Number of clusters}{
		
		Call \emph{get GOS(i$^{th}$ cluster)}
		
		\If{$\Psi$\textsubscript{spread} $>$
			\textcolor{red}{$\Psi_{\textrm{split}}$}}{
			
			Split cluster into two clusters\;
			Assign q\textsubscript{max} and q$_m$ as mean orientations of the
			new clusters \tcp*{Only for initial data sorting}
			Assign atoms from the unsplit cluster to the new clusters by calculating the
			angle-based nearest-neighbor\;	 
			
		}
	}

	\caption{Split clusters}
	\label{algo:split_cluster}
\end{algorithm}

\subsection{Cluster Merge}
Clusters are merged (see \aref{algo:merge_cluster}) based on the inter-cluster disorientation ($\phi$) calculated using the mean orientations of the clusters. \Fref{fig:AlgoSchema}c shows a schematic of two clusters with similar orientations, merged into a single cluster. Any two clusters satisfying the condition, $\phi \ <= \ \phi_{merge}$, with $\phi_{merge}$ being the user-defined merge threshold, qualify as a cluster pair for the merge operation. All such cluster pairs are sorted in ascending order of their disorientation angle ($\phi$) to allow the cluster pairs having the lowest disorientation angle to merge first. Finally, for each of such cluster pair, if neither of the clusters is already merged to other clusters and additionally, if the maximum spread angle ($\Psi$\textsubscript{spread}) of the final cluster after merging is less than the user defined split threshold angle ($\Psi$\textsubscript{split}), they are merged together as a single cluster. This additional condition using the split threshold is to avoid repeated split and merge of the same clusters over the iterations due to the incompatible input threshold values, i.e. split threshold $<<$ merge threshold. %During the final merge clusters call from Algorithm \ref{algo:main_program} line \ref{algo:merge_cluster_pos}, disorientation angles are calculated instead of misorientation angles, considering all 24 cubic symmetries.

%%%%%%%%%%%%%%%%%%%%%%%%%%% REWRITTEN %%%%%%%%%%%%%%%%%%%%%%%%%%%%%%%%%

\newpage
\begin{algorithm}[H]
	\SetAlgoLined
	\KwInput{
		\begin{enumerate}
			
			\item Clusters
			\item \textcolor{red}{$\Phi_{\textrm{merge}}$}
			\item \textcolor{red}{$\Psi_{\textrm{split}}$}
			\item Positional connectivity flag \tcp*{flag = True, only during merge clusters call from Algorithm \ref{algo:main_program} line \ref{algo:merge_cluster_pos}}
			
		\end{enumerate}	
	}              % Set the Input
	\KwOutput{Clusters after merge}
	
	\For{$i\gets1$ \KwTo Number of clusters}{ 
			
			Calculate disorientation angle ($\phi$) between clusters with their mean
			orientations\;}

	Find cluster pairs having $\phi$ $<=$ \textcolor{red}{$\Phi_{\textrm{merge}}$} and sort these pairs based on their disorientation angle in ascending
	order\;

	\For{$i\gets1$ \KwTo Unique cluster pairs}{
		
		\If{\bigg[GOS ($\Psi$\textsubscript{spread}) of the combined cluster after
			merge $<=$ \textcolor{red}{$\Psi_{\textrm{split}}$} \textbf{AND} Both clusters remain
			unmerged in the current merge clusters routine call
			\label{algo:merge_cond2}\bigg]}{
			
			\If{(Positional connectivity flag == True)}{
				
				Check the positional connectivity of the cluster pair\tcp*{To avoid re-merging
					of clusters that were already positionally split}
				
				\If{Connected}{
					
					Merge the cluster pair
					
			}}

			\Else
			{
				
				Merge the cluster pair
				
			}
			
		}
	}

	\caption{Merge clusters}
	\label{algo:merge_cluster}
\end{algorithm}

\newpage
\section{Methods for comparison}
\label{sec:ovito_algo_methods}

To verify the working of our \algo{} algorithm, we compare the results with two other algorithms -- automatic graph clustering and minimum spanning tree -- implemented in the open-source visualization tool, \ovito{} \cite{ovito}. A common feature of both algorithms is that they use a nearest neighbor graph to perform agglomerative hierarchical clustering. These algorithms are available under the \emph{Grain Segmentation Modifier} in \ovito{} and are documented as ``experimental'' versions in the \ovito{} manual \cite{OVITO_grain_seg_manual}. As a result, we only use them for comparing the results of our algorithm for a polycrystalline sample and do not undertake an in-depth study of the algorithms itself. For this study, the options \emph{Adopt orphan atoms} and \emph{Handle coherent interfaces/stacking faults} are enabled. Note that these options are only available with the grain segmentation algorithms of \ovito{}. The former option assigns orphan atoms at, e.g. GBs, to the nearest grain and is similar to the approach used in our \algo{} algorithm. The latter merges atoms having hcp crystal structure with atoms having cubic crystal structures at stacking faults and/or other types of coherent interfaces \cite{OVITO_grain_seg_manual}. This feature is, however, not a part of our \algo{} algorithm.

\subsection{Graph clustering}
The graph clustering algorithm employs the node pair sampling method \cite{bonald2018hierarchical} for building up the grains by contracting edges of a graph. The edge weights of the graph are initialized as 
\begin{equation}\label{eq:WtsGC}
W = \exp\left( - \theta^2/3\right), 
\end{equation}
where $\theta$ is the misorientation angle in degrees between two neighboring atoms. The algorithm contains two clustering parameters - \emph{merge threshold} and \emph{minimum number of atoms per grain}. The \emph{merge threshold}, however, has \emph{no intuitive physical meaning} \cite{OVITO_grain_seg_manual}. The modifier in \ovito{} allows for an automatic mode, where the merge threshold is chosen automatically using a statistical analysis of a sequence of graph merging steps; merging of clusters is stopped as soon as a deviation from the regular exponential behavior is observed \cite{OVITO_grain_seg_manual}. 

\subsection{Minimum spanning tree}
The second algorithm employs misorientation angles between neighboring atoms as the edge weights of a minimum spanning tree. Grains are built up by contracting edges in a sorted order based on misorientation, with the latter also acting as a measure of the merge distance. The approach is similar to that proposed by Panzarino and Rupert \cite{panzarino2014tracking} but uses a hierarchical graph which reduces computation times. The algorithm is fast and has low memory usage. It is pointed out in the \ovito{} manual \cite{OVITO_grain_seg_manual} that this method can lead to poor results in the presence of local perturbations due to, e.g., thermal noise.

\section{Testing and validation}
\label{sec:test_bicrystal}

The \algo{} algorithm is first validated on a bicrystal dataset containing a coherent twin boundary (CTB). For all validations, the maximum number of iterations and the convergence tolerance were set to 200 and 1e-5, respectively. All clustering results presented below are the final results obtained only after achieving convergence to the given tolerance value. Visualization of the data is performed using \ovito{} \cite{ovito}.

\begin{figure}
\centering
\includegraphics[width=\textwidth]{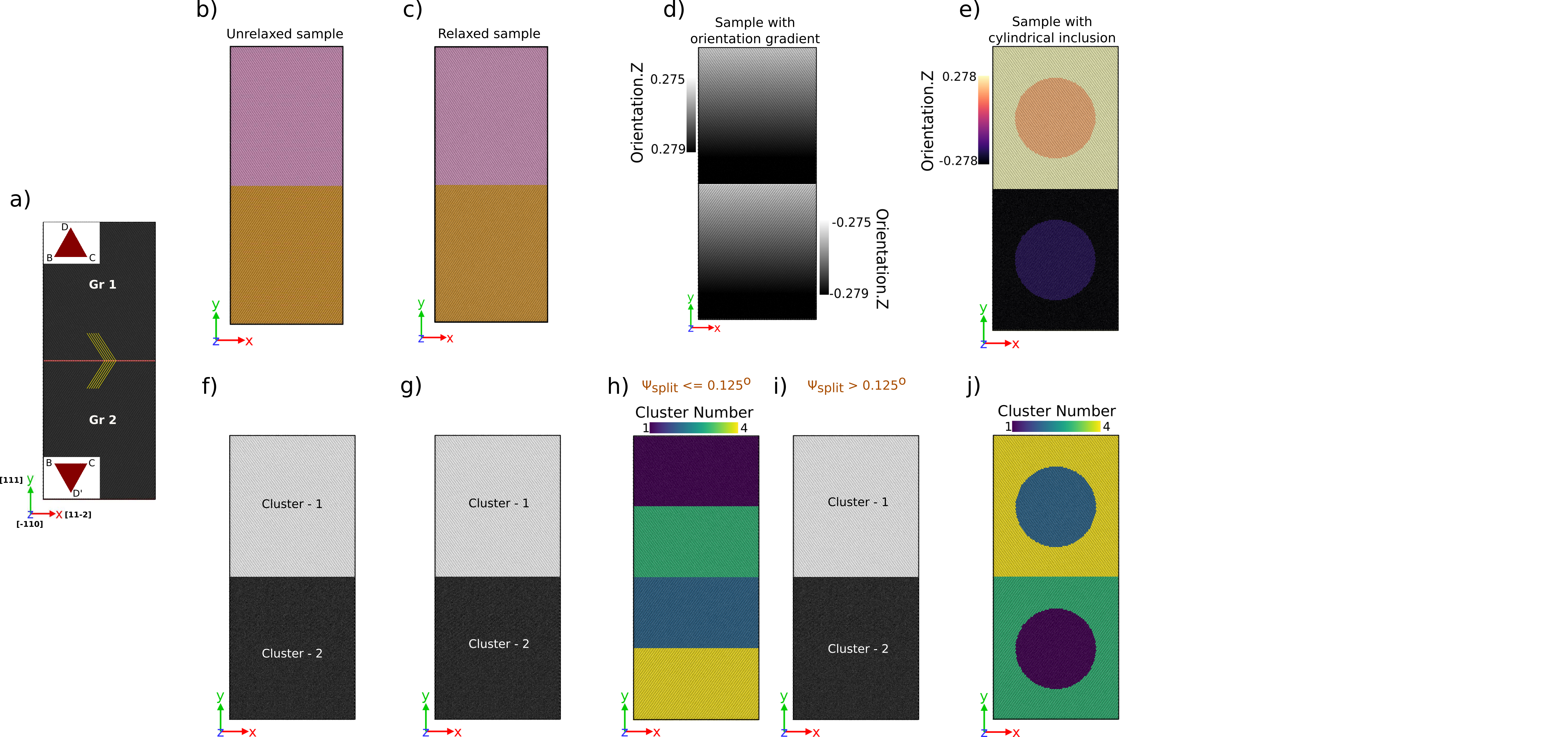}
	\caption{Results of validation and testing of our \algo{} algorithm. a) Au bicrystal sample with a coherent twin boundary and colored based on common neighbor analysis (black: fcc, red: hcp). The orientation of the two grains is shown using the Thompson tetrahedron. The yellow solid lines indicate coherent atomic planes in the two grains. b,c) Unrelaxed and relaxed samples colored based on orientation of individual atoms. d) Synthetic bicrystal sample with an artificial orientation spread in each grain. e) Synthetic bicrystal sample with a cylindrical inclusion with a misorientation of $\approx$~9.5$\degree$ from the matrix. f-j) Clustering results with atoms colored according to their corresponding cluster numbers. 
}\label{fig:bicrystal}
\end{figure}

\Fref{fig:bicrystal}a shows a Au bicrystal with a $\Sigma$3 (111)  CTB constructed using  Atomsk \cite{hirel2015atomsk}. The sample has approximate dimensions of 25 x 56 x 7 nm$^3$ and contains 600,000 atoms. Periodic boundary conditions are applied in all directions. The bicrystal sample is then relaxed using the conjugate gradient and FIRE \cite{guenole2020} minimization algorithms as implemented in the atomistic simulation software LAMMPS \cite{plimpton1995fast,plimpton1997particle}. The interatomic forces are modeled with an embedded atom method (EAM) potential for Au \cite{olsson2010transverse}. \Fref{fig:bicrystal}b,c show the orientations in the unrelaxed and relaxed samples, respectively.

The clustering parameters used for the validation of the \algo{} algorithm, unless specified otherwise, are:
\begin{itemize}
	\item Split threshold angle: $\Psi_{\textrm{split}} = 0.001\degree$
	\item Merge threshold angle: $\Phi_{\textrm{merge}} = 0.001\degree$
	\item Initial number of clusters: $n^{init}_{clus} = 1$ 
	\item Minimum number of atoms per cluster: $D^{min}_g$ = 10
\end{itemize}

The split and merge thresholds are set to very low numerical values since there exists almost no orientation spread in both the unrelaxed as well as the relaxed samples. The initial number of clusters $n^{init}_{clus} = 1$ is chosen so as to avoid any bias in the final result, since the expected number of clusters is known.

Two further synthetic samples, with an artificial orientation spread in each individual grain, are created to further validate the clustering parameters. The first of the synthetic samples (\fref{fig:bicrystal}d) has an orientation spread of $\approx 0.5\degree$ introduced linearly from the GB to the boundary of the simulation box. The second synthetic sample has a central cylindrical region in each individual grain (\fref{fig:bicrystal}e), with a disorientation angle of approximately 9.5$\degree$ to the surrounding matrix.

The results from our \algo{} clustering are shown in \fref{fig:bicrystal}f-j. For the unrelaxed and relaxed cases, the algorithm correctly identifies two clusters delimited by the CTB. For the case with a linear orientation spread, the algorithm identifies four clusters. Increasing the split threshold ($\Psi_{\textrm{split}} $) beyond $\approx 0.125\degree$ results in the two clusters in the top and bottom grain merging together to form a single individual grain. For the sample with a cylindrical inclusion, we obtain as expected, four clusters with the inclusions identified as separate clusters.

\section{Application on a polycrystalline sample}
\label{sec:test_polycrystal}

We demonstrate the working of our \algo{} algorithm on a nanocrystalline thin film sample. The initial structure is generated by means of a constrained Voronoi tessellation \citep{Prakash2017NC,Xu2009} so as to reduce non-equilibrium junctions \cite{serrao2021}. The initial dimensions of the thin film are 180 x 120 x 15 nm$^3$ and contains 122 grains with a mean grain diameter of 15 nm. Each grain is assigned an orientation with \hkl<111> along the film thickness and a random rotation in the plane of the thin film, resulting in purely tilt grain boundaries in the structure. The atomistic structure generated using the open-source toolbox \emph{nano}\textsc{sculpt} \cite{prakash2016b} contains approximately 19~Mio atoms (see \fref{fig:TFundeformedOsISODATA}a). Periodic boundary conditions (PBC) are applied in the plane of the thin film; free boundaries exist along the thickness of the film.

The atomistic sample is then relaxed using the FIRE algorithm in standard molecular statics simulations, and subsequently equilibrated and thermalized at 300~K for 40~ps. \Fref{fig:TFundeformedOsISODATA}b,c shows the defect structure -- computed using common neighbor analysis -- and the orientations of individual grains in the thin film. The equilibrated pressure-free structures are then subjected to uniaxial tension at a constant strain-rate of $10^8 \, s^{-1}$. Simulations are performed with the  ITAP molecular dynamics (IMD) code \cite{stadler1997imd}. The interatomic forces are modeled with an EAM potential for Au \cite{Park2005}.

\subsection{Au undeformed polycrystalline thin film sample at 300 K}

\begin{figure}[htbp!]
\centering
\includegraphics[width=\textwidth]{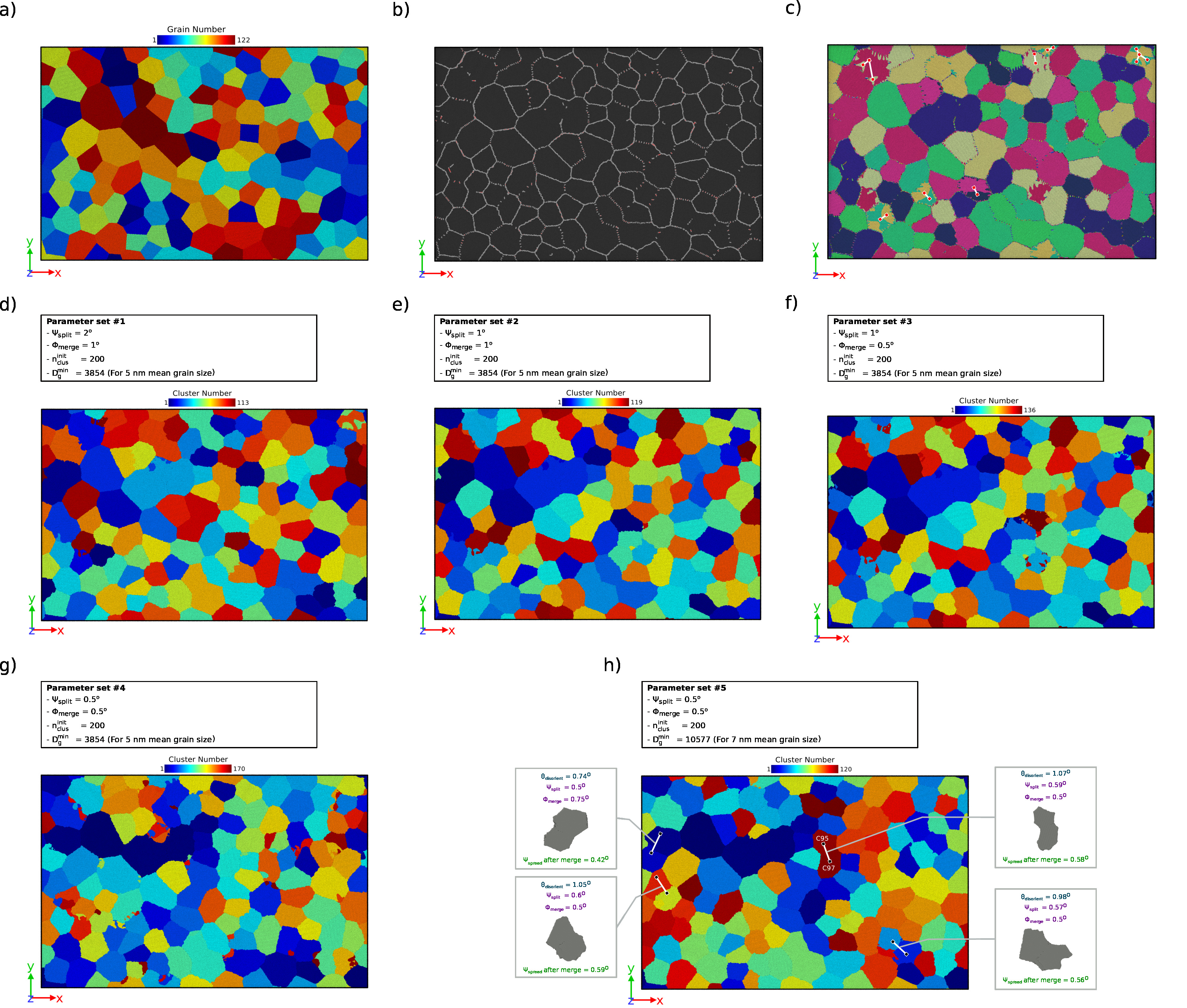}
\caption{Au polycrystalline thin film undeformed sample at 300~K (sliced in the thickness for better visualization). a) Thin film sample with 122 grains. Atoms are colored according to their initial grain numbers. b) CNA results from OVITO (Color code for atoms: black - fcc, red - hcp, grey - others). c) Orientations of individual grains in the thin film sample (in terms of quaternions) of individual atoms are converted and represented in RGB values. The white lines with red marker connect regions with symmetrically equivalent orientations. d-h) Clustering results with five different parameter sets. Atoms are colored according to their cluster numbers. Higher threshold values result in coalescence of certain grain pairs as shown in h).}\label{fig:TFundeformedOsISODATA}
\end{figure}

The results of clustering obtained via the \algo{} algorithm on the undeformed (relaxed + thermalized) sample are shown in \fref{fig:TFundeformedOsISODATA}d-h. The influence of the split and merge threshold values on the final clustering is clearly visible from the results. As the threshold values decrease, the number of clusters increases. With parameter set~\#1, the \algo{} algorithm is able to identify clusters which conform approximately to the orientation map in \fref{fig:TFundeformedOsISODATA}c. For $\Phi_{\textrm{merge}} \leq 1\degree$ (see \fref{fig:TFundeformedOsISODATA}e-g), small clusters emerge at GBs and triple junctions (TJ), where the local orientation of individual atoms differs from the mean orientation of the grains.

Increasing the threshold for the minimum number of atoms per cluster to an equivalent grain size of 7 nm ensures that such small clusters are now treated initially as orphan clusters and assigned to the closest cluster after the final iteration. Nonetheless, grains separated by very low angle GBs ($<1\degree$) can still be identified, as shown exemplarily by the highlighted grains in \fref{fig:TFundeformedOsISODATA}h. Using threshold values beyond those defined by the orientations of such grains results in the coalescence of such grain pairs to a single cluster, see highlighted boxes in \fref{fig:TFundeformedOsISODATA}h.

\begin{figure}[htbp!]
\centering
\includegraphics[height=0.9\textheight]{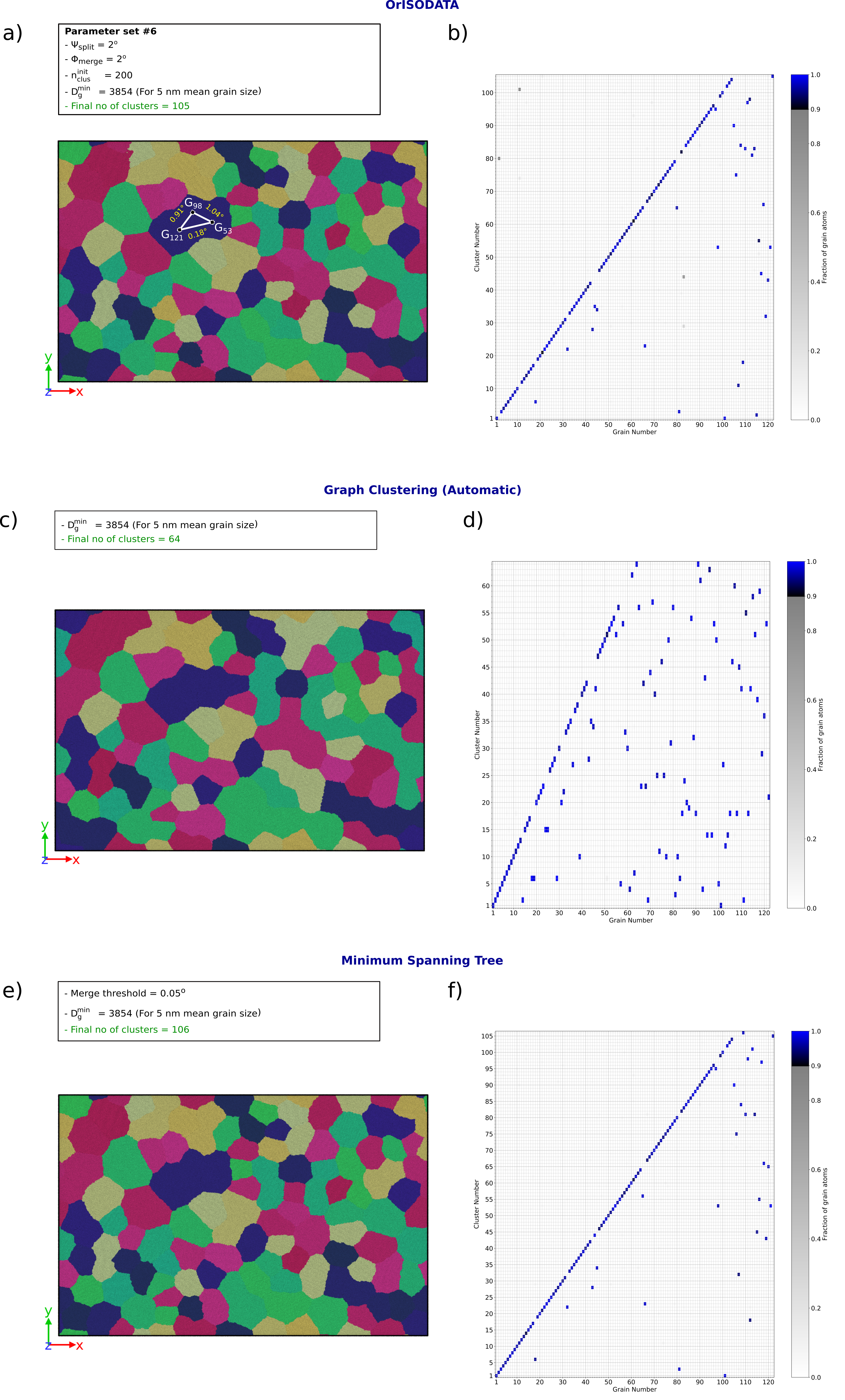}
\caption{Comparison of clustering results from \algo{} algorithm with those from grain segmentation algorithms in \ovito{}. \emph{left}: Individual clusters colored by their mean orientations by mapping the corresponding quaternion ($q_x,q_y,q_z,q_w$) to red,green and blue values, respectively. \emph{right}: Correlation plot between initial grain number and the final cluster identified by the corresponding algorithm.}\label{fig:TFUndefCompareClusters}
\end{figure}

Since we do not expect the orientation of individual grains to change as a result of relaxation and equilibration of the sample, atoms forming the original grain must be assigned to a new distinct cluster, as long as the threshold criteria are satisfied. This is evident in \fref{fig:TFUndefCompareClusters}b which shows the fraction of atoms of each grain assigned to different clusters, thus providing a measure of effectiveness of the algorithm. Such a plot is particularly helpful to track, e.g., whether the initial grains are split or merged with the neighboring grains whilst being assigned to a cluster. The diagonal trend indicates that by-and-large each grain is assigned uniquely to a new cluster. Multiple points along a single horizontal line indicate the coalescence of multiple grains into a single cluster. This is for instance the case with grains 53, 98, 121 which are now assigned to the cluster 53. Such merging of two or more grains is a result of very low misorientation between the grains, which is less than the threshold of $2\degree$; the GBs between such grains contain no defect atoms as seen in \fref{fig:TFundeformedOsISODATA}c. On the other hand multiple points along a single vertical line indicate that atoms in the original grain are split into multiple clusters as a consequence of local rearrangement close to GBs. Only three grains evidence a split where more than 10\% of the original grain is assigned to a neighboring cluster.

The results of clustering with the grain segmentation algorithms in \ovito{} are shown in \fref{fig:TFUndefCompareClusters}c--f. The automatic graph clustering (AGC) algorithm results in fewer clusters (64) than with our \algo{} algorithm (cf. \fref{fig:TFUndefCompareClusters}c). In particular, grains separated by low angle GBs are organized to a single cluster with as many as five grains sometimes grouped into a single cluster. Hence a significant deviation from a purely diagonal trend is seen in the correlation plot shown in \fref{fig:TFUndefCompareClusters}d. The threshold value determined by the AGC algorithm is 22 units, which results in a misorientation angle of approximately $\approx 3\degree$ using \eref{eq:WtsGC}. Increasing the threshold value in \algo{} to $3\degree$ leads, expectedly, to fewer number of 98 clusters, but is still much higher than that obtained with the AGC algorithm (see \emph{supplementary} \fref{SuppFig:OrisodataSameThresAGC}).

%\todo{More info needs to be provided/discussed here. For example, is a change from 2 to 3$\degree$ in misorientation the reason for this huge change?}

By contrast, with the minimum spanning tree (MST) algorithm \fref{fig:TFUndefCompareClusters}e we obtain an almost equal number of clusters as with our \algo{} algorithm -- 106 wth MST vs. 105 with \algo{}. The clustering pattern obtained from the MST algorithm is very similar to that observed from our \algo{} algorithm. The low merge threshold of $0.05\degree$ is seemingly the reason behind the ability to recognize low angle GBs that are apparently neglected with the AGC method. This threshold value is, however, not directly comparable with the merge threshold in the \algo{} algorithm since the latter is a measure of distance between the mean orientations of two grains, whereas the former defines a threshold for the misorientation between neighboring atoms.

Despite these noticeable differences in the clustering results from the three algorithms, no large differences can be seen in the mean orientations of the identified clusters. The pole figures shown in \emph{supplementary} \fref{SuppFig:StereoProjUndef} are very similar and show the presence of a \hkl{111} texture in the thin film.

\subsection{Au deformed polycrystalline thin film sample at 300 K}

\begin{figure}[htbp!]
\centering
\includegraphics[height=0.8\textheight]{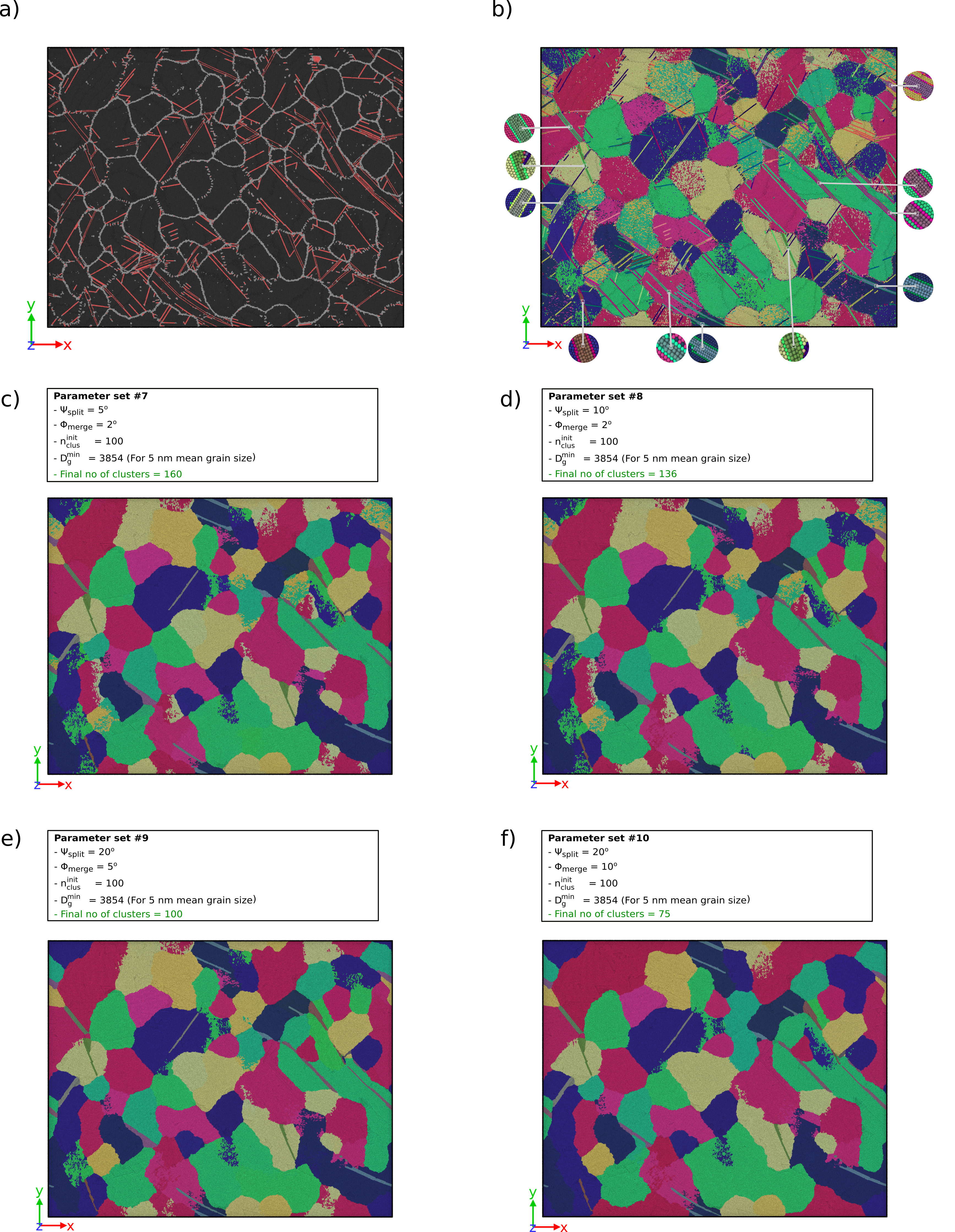}
\caption{Au deformed polycrystalline thin film sample (Sample has been sliced in the Z direction for better visualization). a) CNA results from \ovito{} (Color code for atoms: black - fcc, red - hcp, grey - others). b) Orientations (in terms of quaternions) of individual atoms are converted and represented in RGB values. Some of the stacking fault regions present within the grains are highlighted. c-f) Clustering results with four different parameter sets. Atoms are colored according to their mean orientations. }\label{fig:TFdefOrISODATA}
\end{figure}

The defect structure in the thin film sample, after 10\% tensile strain in the global y-direction, obtained via CNA is shown in \fref{fig:TFdefOrISODATA}a. The deformed configuration is dominated primarily by stacking faults and deformation twins, which form via the motion of partial dislocations. Very few full dislocations are visible. The orientation map (see \fref{fig:TFdefOrISODATA}b) shows a larger spread of orientations in many grains, than that seen in the undeformed configuration. Twinned regions are visible as regions with a different orientation in comparison to the orientation of the parent grain. It must be expected that a good grain segmentation algorithm is able to identify such regions and help track the evolution of, e.g. twin volume during deformation.

\begin{figure}[htbp!]
\centering
\includegraphics[height=0.9\textheight]{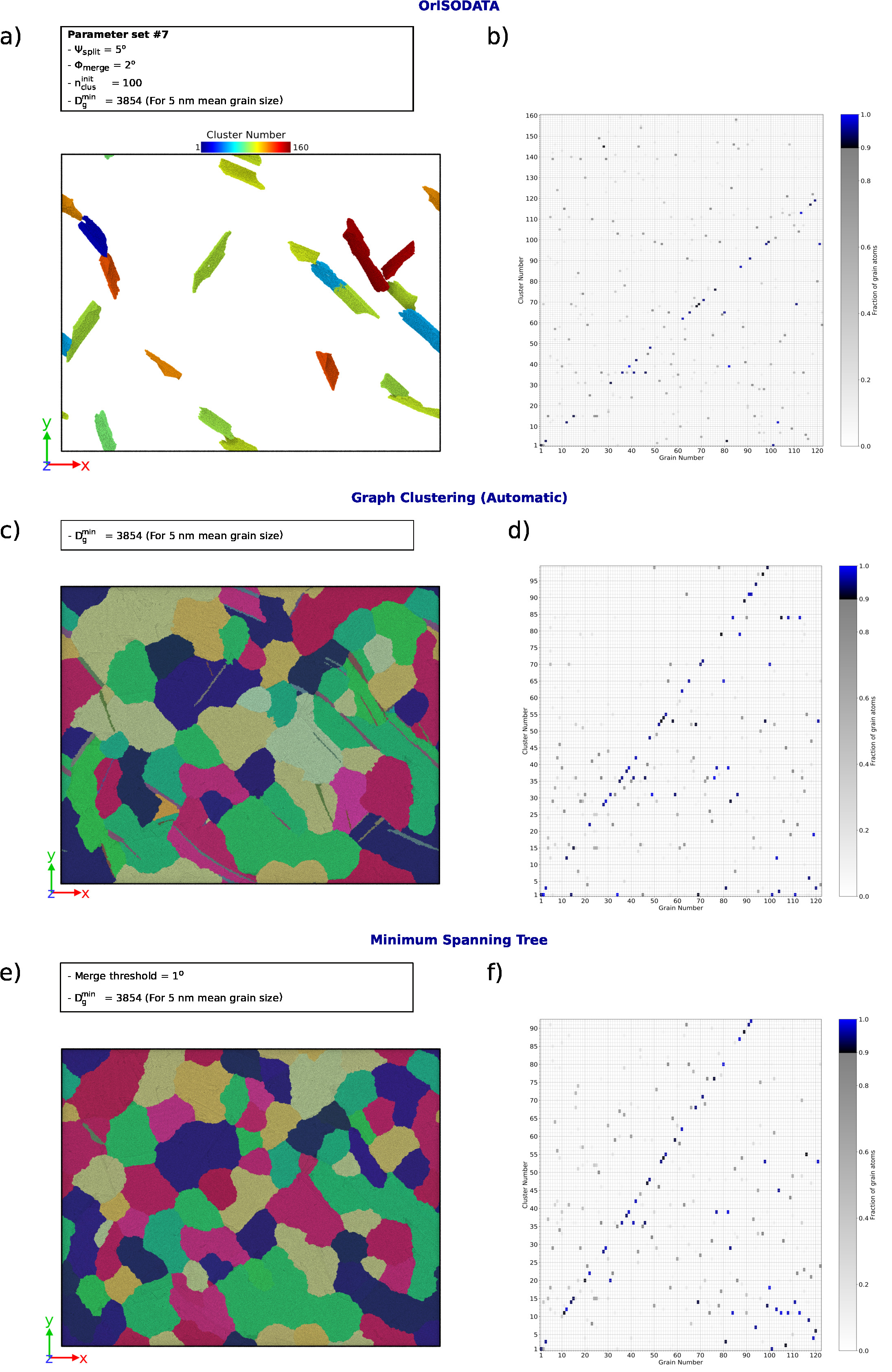}
\caption{Comparison of clustering results from \algo{} algorithm with those from grain segmentation algorithms in \ovito{} on the Au polycrystalline sample deformed to 10\% tensile strain. \emph{left}: Individual clusters colored by their corresponding mean orientation. \emph{right}: Correlation plot between initial grain number and the final cluster identified by the corresponding algorithm.} 
\label{fig:TFdefCompareClusters}
\end{figure}

Comparison of clustering results from \algo{} with those from the grain segmentation algorithms in \ovito{} throws up some interesting observations, see \fref{fig:TFdefCompareClusters}. The number of clusters identified by \algo{} is significantly higher than the other two algorithms. A purely diagonal trend in the correlation between the undeformed cluster/grain and the identified cluster (\fref{fig:TFdefCompareClusters}b,d,f) is no longer visible in any of the algorithms. This observation can be ascribed to the larger spread of orientation gradients in many grains, as a result of which, atoms belonging to individual undeformed grains/clusters are split into multiple clusters in the deformed state. Furthermore, many clusters have now formed via the conglomeration of multiple initial grains, indicating significant changes in GB characteristics, topology and network. Of the two algorithms in \ovito, the AGC method performs better than the minimum spanning tree approach in identifying local twinned regions. We note that the option \emph{Handle coherent interfaces/stacking faults} is switched on during clustering with the grain clustering algorithms in \ovito{}. Turning this option off results in improved identification of the twinned regions with the AGC method (see \emph{supplementary} \fref{SuppFig:SFflag}). No change in the identification of twins is seen with the MST algorithm, but switching off the option results in substantially more clusters (see \emph{supplementary} \fref{SuppFig:SFflag}).

The mean orientations of individual clusters confirm the aforementioned observations. With the MST algorithm, almost no change is observed in the overall texture (cf. \emph{supplementary} \fref{SuppFig:StereoProjDef}). By comparison, a strong deviation from the strong initial \hkl{111} texture is observed in a few clusters (mostly corresponding to deformation twins) with both \algo{} and the AGC algorithms. The few points which deviate from the initial \hkl{111} texture with the MST algorithm correspond to a single twin identified by the algorithm.

%\todo{Stereographic projection. Mention differences to the texture of the unrelaxed sample, and how the texture of clusters from min. spanning tree is similar to that of the undeformed sample (no changes in texture), whilst spread of points can be seen in the other two algorithms. Do these relate to twins?}

\section{Discussion}
\label{sec:discussion}

The primary difference between the grain segmentation algorithms in \ovito{} and our \algo{}  algorithm is in the approach taken. With the graph based methods implemented in \ovito{}, a more-or-less bottom-up approach is forged, wherein clustering begins by the creation of a graph representation of the input structure using atoms as the graph nodes and the bonds with neighbors as the graph edges. If the threshold criterion is satisfied, graph edges are collapsed resulting in a hierarchical clustering of individual grains. By contrast, in the \algo{} algorithm we follow a top-down approach. Clustering begins directly in the data space without reference to individual atoms. Clusters are then split and merged if certain criteria are met. The reference to individual atoms appears only now: If a cluster has multiple distinct regions which are not connected to each other in terms of immediate neighbors, we split such clusters further.

\begin{figure}[htbp!]
	\centering
	\includegraphics[width=1\textwidth]{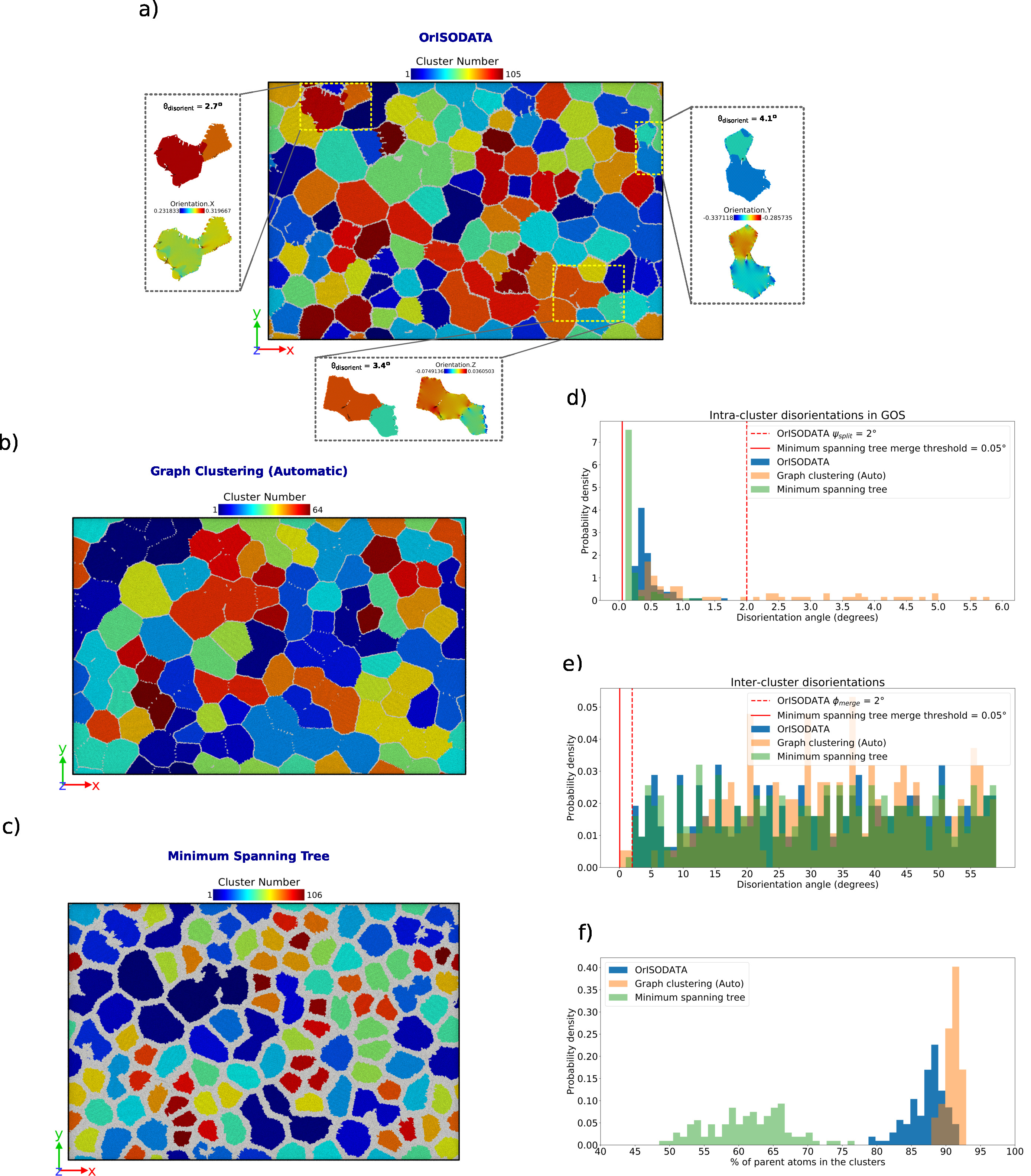}
	\caption{Comparison of clustering results on the undeformed sample of \algo{} algorithm with the grain clustering algorithms in OVITO. The samples are sliced in the thickness for better visualization. a) Clustering results from \algo{} for the parameter set \#6. Disorientation angles and the color map showing the quaternion component having maximum variation between the sub-clusters formed within grains, are highlighted separately. b) Clustering results from graph clustering (automatic). c) Clustering results from minimum spanning tree algorithm. d) Intra cluster disorientation. e) Disorientation between neighboring clusters. f) Distribution of percentage of parent atoms (non-orphan atoms) in individual clusters. The threshold values used for the corresponding algorithms are marked in vertical lines. Color code in a-c) - Atoms are colored according to their cluster numbers except the orphan atoms which are colored in light gray.}
	\label{fig:CompareCharacteristicsUndeformed}
\end{figure}

\begin{figure}[htbp!]
\centering
\includegraphics[width=\textwidth]{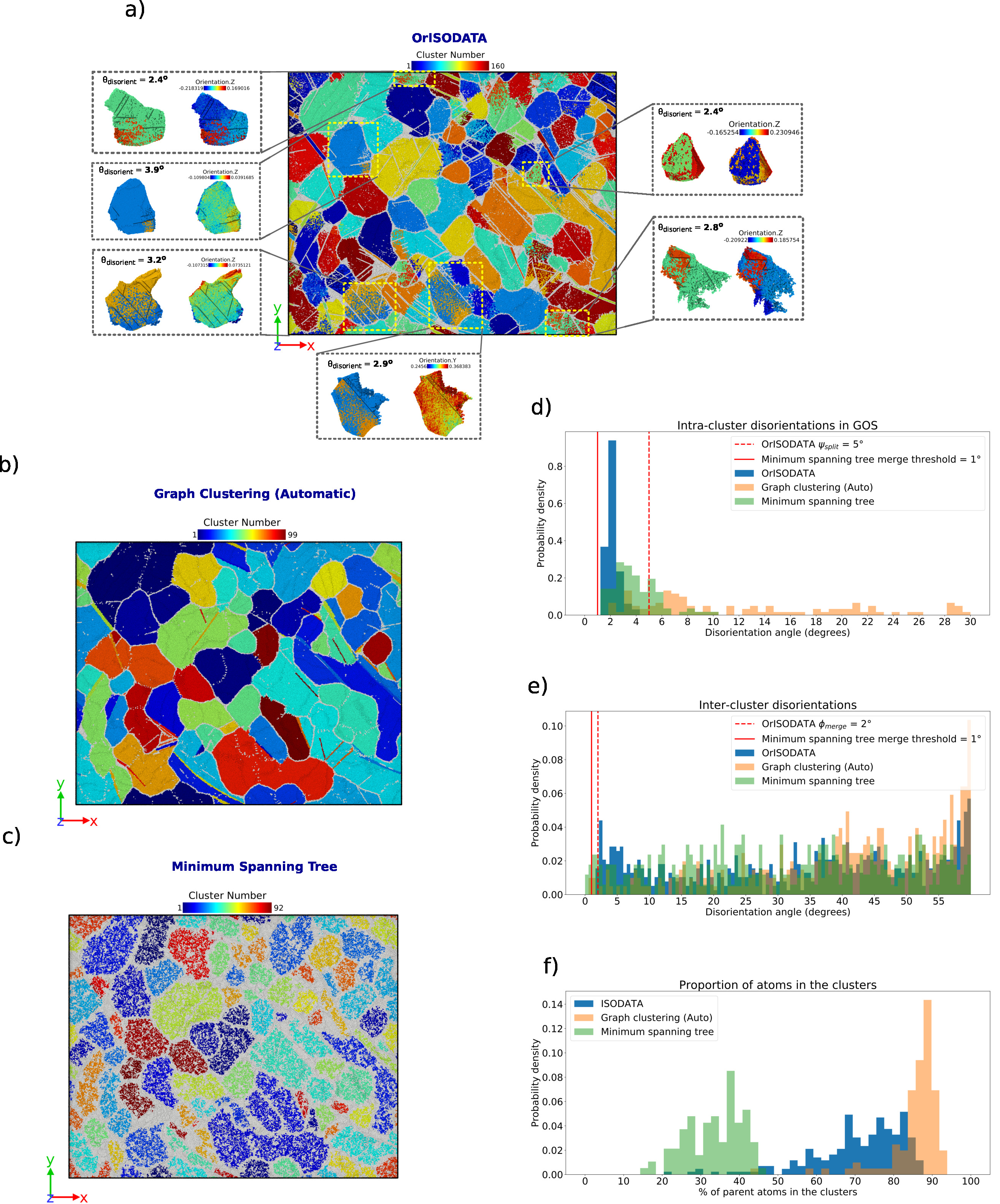}
\caption{Comparison of clustering characteristics from \algo{}, graph clustering and minimum spanning tree algorithms in the strained sample. The samples are sliced in the thickness for better visualization. a-c) Clusters determined by the three different algorithms before the adoption of orphan atoms. Cluster numbers for the parameter set \#7 is shown in a). d) Intra cluster disorientation. e) Disorientation between neighboring clusters. f) Distribution of percentage of parent atoms (non-orphan atoms) in individual clusters.  The threshold values used for the corresponding algorithms are marked in vertical lines. Color code in a-c) - Atoms are colored according to their cluster numbers except the unadopted atoms which are colored in light gray.}
\label{fig:CompareCharacteristicsDeformed}
\end{figure}

Despite the different approaches taken, the working of the algorithms can be understood by comparing the characteristics of the output clusters and evaluating the influence of the corresponding parameters on the clustering. One aspect of the clustering process is the treatment of orphan atoms. \Fref{fig:CompareCharacteristicsUndeformed} and \fref{fig:CompareCharacteristicsDeformed} show the clustered configurations of the undeformed and deformed samples, respectively, from the three algorithms, before the adoption of orphan atoms. It is evident that for the parameter sets chosen, the fewest number of orphan atoms is with the AGC method and the highest with the MST algorithm. Approximately 35\% of the undeformed sample and 65\% of the deformed sample is identified as orphan atoms with the MST algorithm (see \fref{fig:CompareCharacteristicsUndeformed}f and \fref{fig:CompareCharacteristicsDeformed}f). The number of orphan atoms with the \algo{} algorithm is close to that of the AGC method; the slight differences seen in the distributions are ascribed to the chosen parameter set which results in many more clusters being identified with our \algo{} algorithm in comparison to the AGC method.

This increased fraction of orphan atoms is directly responsible for the poor identification of deformation twins with the MST algorithm. With all the three algorithms, stacking faults are identified as orphan atoms due to their hcp crystal structure. However, with MST, regions around hcp atoms are also identified as orphan atoms. These orphan atoms are later assigned to a parent cluster, resulting in twinned regions appearing much larger or being completely assimilated into the parent grain.

The threshold parameter used in the MST algorithm evidences a high-sensitivity to small perturbations. Choosing the right threshold parameter in the MST algorithm can hence be quite tricky, as seen in \emph{supplementary} \fref{SuppFig:MergeThresholdMST}. For the deformed configuration, the chosen value of $1\degree$ results in the best clustering pattern that is comparable with the other algorithms, as shown in \fref{fig:TFdefCompareClusters}. Decreasing the threshold to $0.9\degree$ results in almost equal number of clusters, but significantly different clustering pattern. Increasing the threshold to $1.1\degree$ results in a higher number of clusters, see \emph{supplementary} \fref{SuppFig:MergeThresholdMST}. This behavior is non-intuitive, since we expect the number of clusters to either increase or remain constant with a decrease in the threshold value, and can again be attributed to the presence of large number of orphan atoms with the MST algorithm. Decreasing the minimum number of atoms per cluster to 1000 reduces the number of orphan atoms and results in the expected behavior of increased number of clusters with a lower threshold. The predicted clusters, however, are quite different to those observed with the AGC and \algo{} algorithms.

A quantitative comparison of the three algorithms can be made by looking into the intra-cluster disorientation, and the inter-cluster disorientation between immediate neighbors of a cluster. The results of such a comparison is shown in \fref{fig:CompareCharacteristicsUndeformed} and \fref{fig:CompareCharacteristicsDeformed} for the undeformed and deformed samples, respectively. With MST and \algo{} algorithms where criteria can be specified, such threshold criteria are well satisfied. The intra cluster disorientation in terms of GOS is higher than the threshold of 0.05$\degree$ between atoms in the case of the MST algorithm in the undeformed configuration. This seemingly translates to a maximum GOS of $1.5\degree$ and an average value of $0.1\degree$ in the formed clusters. With the AGC algorithm, the maximum GOS observed was close to $6\degree$. 

In the deformed configuration, a threshold of $1\degree$ with the MST algorithm results in a maximum intra-cluster disorientation of roughly $10\degree$. In comparison, the maximum GOS is $\approx 30\degree$ with the AGC model, which is, however, an artifact of hcp atoms identified as part of the parent cluster. The inter-cluster disorientations evidence a distribution over a range of angles, similar to that seen in the undeformed configuration; both \algo{} and AGC algorithms result in higher number of clusters with large inter-cluster disorientations due to their ability to identify twinned regions. The \algo{} algorithm is, furthermore, clearly able to identify clusters with low angle inter-cluster disorientations, whereas very few of such clusters were identified with the AGC model (see \fref{fig:CompareCharacteristicsDeformed}e).

Using the manual version of the grain clustering algorithm can provide the user with increased flexibility in identifying clusters. Decreasing the threshold value results, as expected, in increased number of clusters (see \emph{supplementary} \fref{SuppFig:ManualGC}). However, the non-physical nature of the threshold used to collapse graph nodes essentially hinders robust interpretation and verification of results. With the undeformed configuration, a threshold of 22 units was used to generate the clustering results shown in \fref{fig:TFUndefCompareClusters}c, which translates to a disorientation angle of approximately 3$\degree$, suggesting a minimum inter-cluster disorientation of the same measure. By fine tuning the parameters in the \algo{} algorithm, we were able to drastically reduce the number of clusters to match that of the graph clustering algorithm (see \emph{supplementary} \fref{SuppFig:AGCcompareOrisodata}). A similar clustering pattern was only obtained for split and merge thresholds of $10\degree$ with our \algo{}. Furthermore, with the graph clustering algorithm, whilst clusters with intra-cluster disorientations of as high as $\approx 4\degree$ are identified as a single cluster, clusters with lower intra-cluster disorientation of $\approx 3.2$ are split into two clusters whose inter-cluster disorientation is a mere $1.2 \degree$. This essentially leads us to conclude that the threshold parameter in the graph clustering algorithm, and consequently, the clustering results, lack simple interpretation.

%for the threshold value of 22 units in the AGC algorithm used for determining the clustering pattern in the undeformed configuration, we obtain initial graph weights of approximately 3$\degree$, suggesting a minimum inter-cluster misorientation of the same measure. 

The clustering results with our \algo{} algorithm depend strongly on the two threshold parameters -- higher numerical values of these thresholds result in fewer number of final clusters, and vice versa. The choice of numerical values for the thresholds depends, however, on the sample and application at hand. For example, to identify low angle GBs, the \emph{merge threshold} must be set to a low value that corresponds to the disorientation between two neighboring grains. On the other hand, the \emph{split threshold} influences the position of a GB in a contiguous domain with significant orientation gradients. In other words, the intra-cluster disorientation must be less than the split threshold and the inter-cluster disorientation between nearest neighbors must be greater than the merge threshold. In both the undeformed and deformed configurations, the results clearly follow this rule, see \fref{fig:CompareCharacteristicsUndeformed}d,e and \fref{fig:CompareCharacteristicsDeformed}d,e.

Two further parameters form a part of the \algo{} algorithm: a) Minimum number of atoms or minimum grain size ($D^{min}_g$) for any cluster to be accepted as a possible solution, b) Initial number of clusters ($n^{init}_{clus}$). The former has an effect on the fraction of orphan atoms, and eliminates the formation of very small clusters (see \fref{fig:TFundeformedOsISODATA}g,h). Nevertheless, for the same threshold parameters, the influence of the minimum grain size on the clustering results is negligible. The influence of the latter parameter, $n^{init}_{clus}$, depends on the orientation gradients in the sample -- larger the gradients in orientation, larger is the influence on the final number of clusters, as is the case with the deformed thin film sample (see \emph{supplementary} \fref{SuppFig:InfluInitClus}). Nevertheless, this variation is less than $\pm 5\%$ indicating the robustness of the entire procedure.

%For the bicrystal sample without orientation gradients, varying the initial number of clusters has no influence on the final outcome (see \fref{SuppFig:InfluInitClus}). For the undeformed thin film sample with minimal orientation gradients, only a small variation in the final number of clusters is seen over all realizations of $n^{init}_{clus}$. For the deformed thin film sample with significant orientation gradients, a slightly larger variation of roughly 5\% in the final number of clusters is seen. 

The advantage of the \algo{} algorithm over the other two algorithms is in the intuitive nature of the threshold parameters which is expressed as follows:
\begin{itemize}
\item \emph{Split threshold}: How much of a deviation from the mean orientation is to be allowed within the domain of the grain
\item \emph{Merge threshold}: How close in terms of orientation should two neighboring grains/clusters be in order for them to be treated as a single grain
\end{itemize}
This intuitiveness of the parameters ensures easy verification and interpretation of results. Furthermore, these threshold parameters are similar in interpretation to those used in experiments (e.g. EBSD) making the methodology appealing to a larger user base and facilitating increased synergy between simulations and experiments.

\section{Conclusions}

In this work, we propose an orientation-based iterative self organizing data analysis (\algo{}) algorithm to identify grains in atomistic simulations of polycrystalline materials. The algorithm takes a data based approach to cluster points in the orientation space. We first validate the algorithm on a bicrystal sample containing a coherent twin boundary. Subsequently, the working of the \algo{} is demonstrated on a polycrystalline thin film sample in the undeformed state as well as the deformed state after 10\% tensile strain at 300~K. Furthermore, the results from our \algo{} algorithm are compared with those from the automatic graph clustering and minimum spanning tree algorithms implemented in \ovito{}. With the latter, we are able to identify very low angle GBs which was not possible with the AGC algorithm. On the other hand, deformation twins are not identified by the MST algorithm, but were clearly identified by the AGC algorithm. 

The findings of the current study can be summarized as follows:
\begin{itemize}
\item The \algo{} algorithm uses parameters which are intuitive and physically relevant for the application. This makes it substantially easy to choose optimal parameters, and to validate and interpret the clustering results.
\item Very low angle grain boundaries, i.e. inter-cluster disorientation less than $2\degree$ can be identified with both the \algo{} and the MST algorithms, but not with the AGC algorithm.
\item Twinned regions in the deformed polycrystalline thin film are identified with the \algo{} and AGC algorithms, but not with the MST algorithm.
\item The proportion of atoms identified as \emph{orphan atoms} is a significant influencing factor on the clustering results of the three algorithms. Perturbations due to the applied strain and local elastic fields can result in a significant fraction of atoms -- 60$\sim$70\% fraction -- in the deformed thin film being classified as orphan atoms with the MST algorithm.
\item A smaller fraction of orphan atoms is the key to proper identification of twinned regions in the deformed state of the polycrystal. Thin twinned regions are otherwise classified as orphan atoms and are later assigned to the nearest cluster.
\item Choosing the right parameter set for the usage of the AGC algorithm is particularly difficult since the threshold parameter has no intuitive meaning. Furthermore, intra-cluster grain orientation spread and inter-cluster disorientation distributions are also not along expected lines.  

\end{itemize}

The two threshold parameters in the \algo{} algorithm are intuitive and relate to similar thresholds used in experiments: the \emph{split threshold} is a measure of the maximum orientation spread within a grain, whereas the \emph{merge threshold} is a measure of proximity between the mean orientation of two grains. We conclude that although the choice of the parameter values strongly influences the clustering results, as is the case with any algorithm, having model parameters that are intuitive and physically relevant to the application, is substantially helpful for choosing  optimal parameter values and also for analyzing and rightly interpreting the final results.

%\newpage
\section*{Acknowledgments}
SS acknowledges funding from the ERC starting grant, ``A Multiscale Dislocation Language for Data-Driven Materials Science'', ERC Grant agreement No. 759419 MuDiLingo. The authors wish to acknowledge the Centre for Information Services and High Performance Computing [Zentrum für Informationsdienste und Hochleistungsrechnen (ZIH)], TU Dresden for providing the computing time for molecular dynamics simulations in the project \emph{NCthinFilms}. 

\section*{References}
\bibliography{Grain_identification_using_isodata_references}

\newpage
\section*{Supplementary material}
\setcounter{section}{0}
\setcounter{figure}{0}
\setcounter{equation}{0}
\renewcommand{\figurename}{Supplementary Figure}
\renewcommand\thefigure{S\arabic{figure}}

\section{Misorientation angle calculation using orientation distance metric}

The orientation distance metric used in the current work is the geodesic distance metric \cite{hartley2013rotation,huynh2009metrics} using quaternions. The quaternions $q$, are represented in the format (q$_x$, q$_y$, q$_z$, q$_w$) having the scalar (q$_w$) at the last. 

With quaternions $q_A$ (q$^A_x$, q$^A_y$, q$^A_z$, q$^A_w$) and $q_B$ (q$^B_x$, q$^B_y$, q$^B_z$, q$^B_w$), the dot product between them is given by,
\begin{equation}
d \ = \ q^A_x.q^B_x \ + \ q^A_y.q^B_y \ + \ q^A_z.q^B_z \ + \ q^A_w.q^B_w
\end{equation}

Using the scalar dot product result (d), the misorientation angle in radians, which is the orientation distance, is given by,
\begin{equation}
\theta \ = \ 2 cos^{-1} (d)
\end{equation}

\newpage

\begin{figure}[htbp!]
\centering
\includegraphics[width=\textwidth]{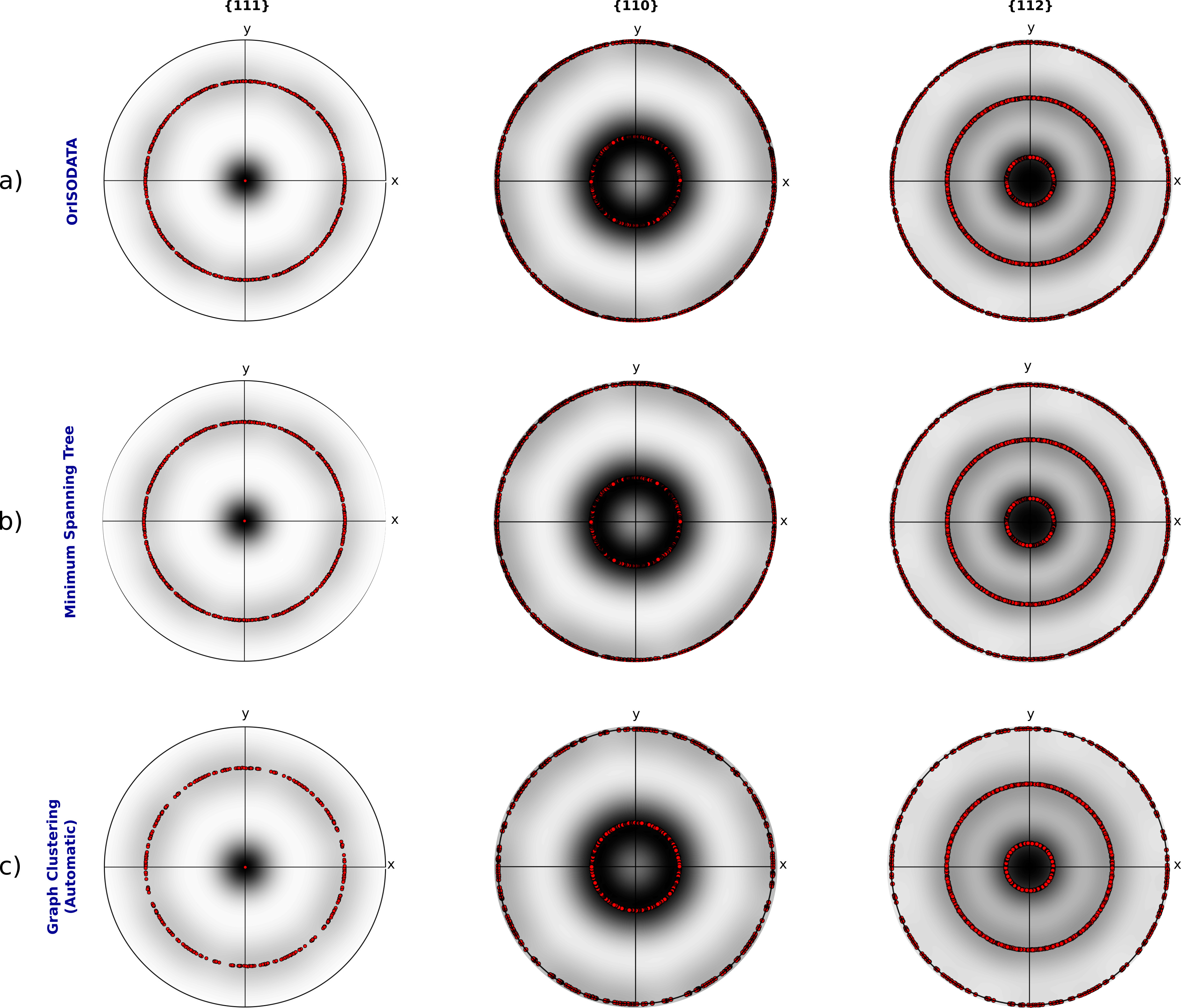}
\caption{Stereographic projections of the mean orientation of individual clusters in the undeformed Au thin film, identified by a) \algo{}, b) Minimum spanning tree, and c) Graph clustering (automatic). The mean orientations are calculated using only the parent atoms, i.e. without the orphan atoms.}
\label{SuppFig:StereoProjUndef}
\end{figure}

\newpage
\section{Additional grain segmentation results from OVITO}

\begin{figure}[htbp!]
\centering
\includegraphics[width=\textwidth]{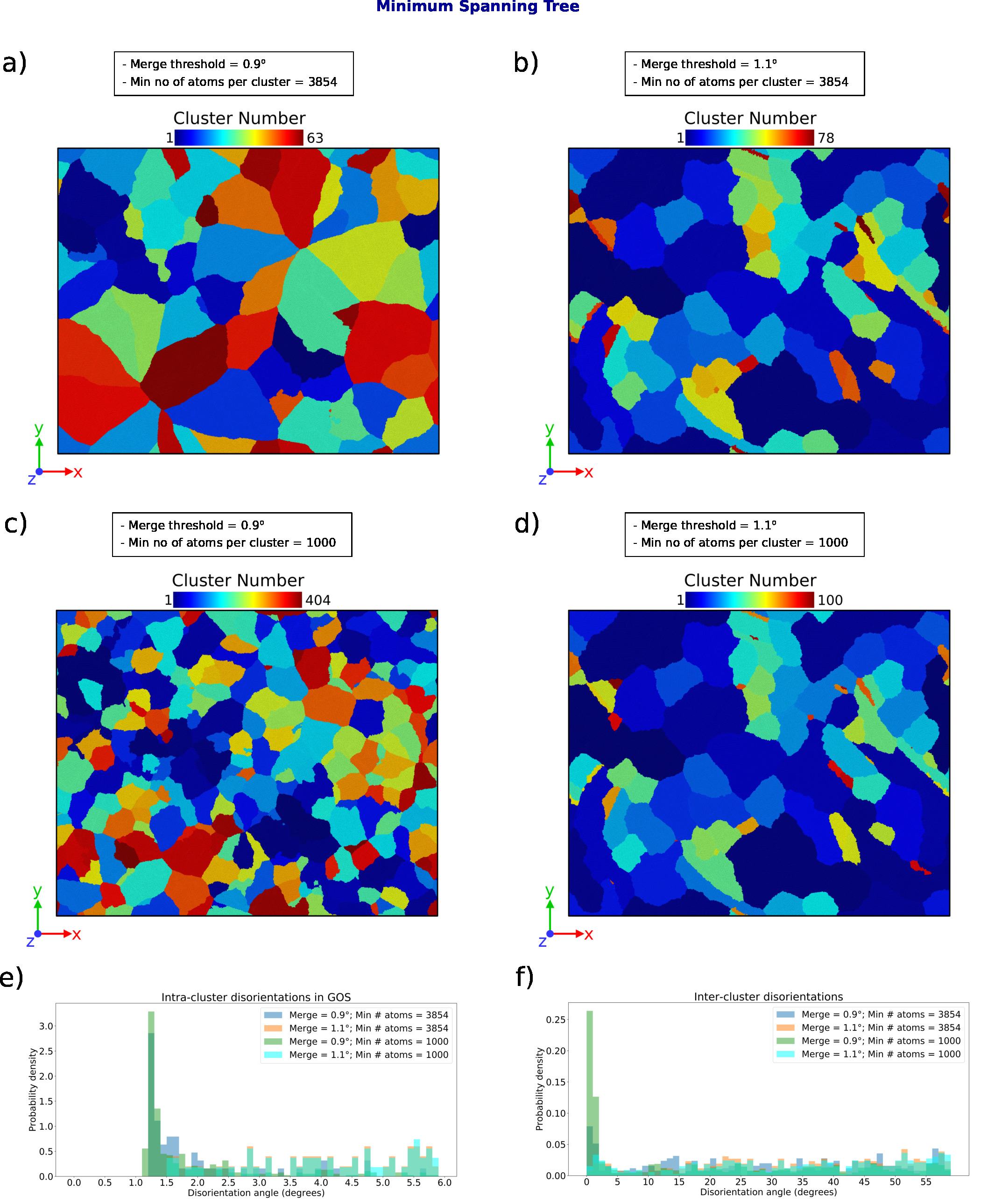}
\caption{Influence of the merge threshold on the clustering results with the minimum spanning tree algorithm.}
\label{SuppFig:MergeThresholdMST}
\end{figure}

This section contains the clustering results from \ovito{} for different merge threshold values and the results without enabling the \emph{Handle coherent interfaces/stacking faults} feature for the deformed 122 grains polycrystalline thin film sample case. \Fref{SuppFig:MergeThresholdMST} a,b), shows the influence of using merge threshold lower and higher than 1$\degree$ on the final number of clusters. Fewer number of final clusters are obtained for both the 0.9$\degree$ and 1.1$\degree$ cases using 3854 atoms as the minimum number of atoms per cluster. By lowering the minimum number of atoms per cluster shown in \fref{SuppFig:MergeThresholdMST} c,d), the number of small clusters have blown up particularly for the 0.9$\degree$ case.

\begin{figure}[htbp!]
\centering
\includegraphics[width=\textwidth]{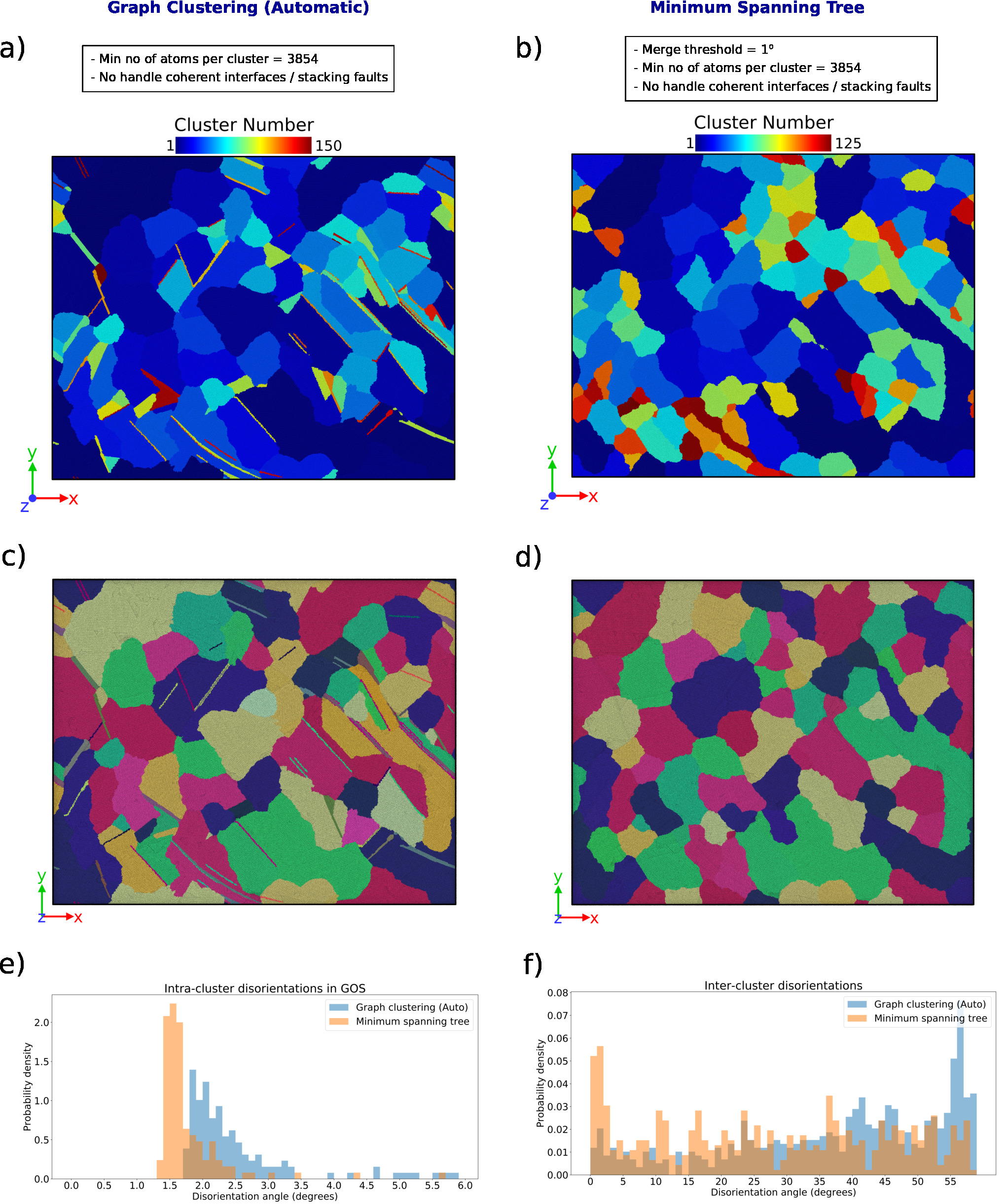}
\caption{Influence of switching off \emph{Handle Coherent Interfaces/Stacking Faults} flag on the clustering results of the automatic graph clustering and the minimum spanning tree algorithms}
\label{SuppFig:SFflag}
\end{figure}

\Fref{SuppFig:SFflag} shows the clustering results
without enabling the \emph{Handle coherent interfaces / stacking faults} feature
for \ovito{} algorithms. The automatic graph clustering algorithm shows additional stacking
fault regions as individual clusters. For minimum spanning tree algorithm, no
significant stacking fault regions are observed after disabling this feature.

\begin{figure}[htbp!]
\centering
\includegraphics[width=\textwidth]{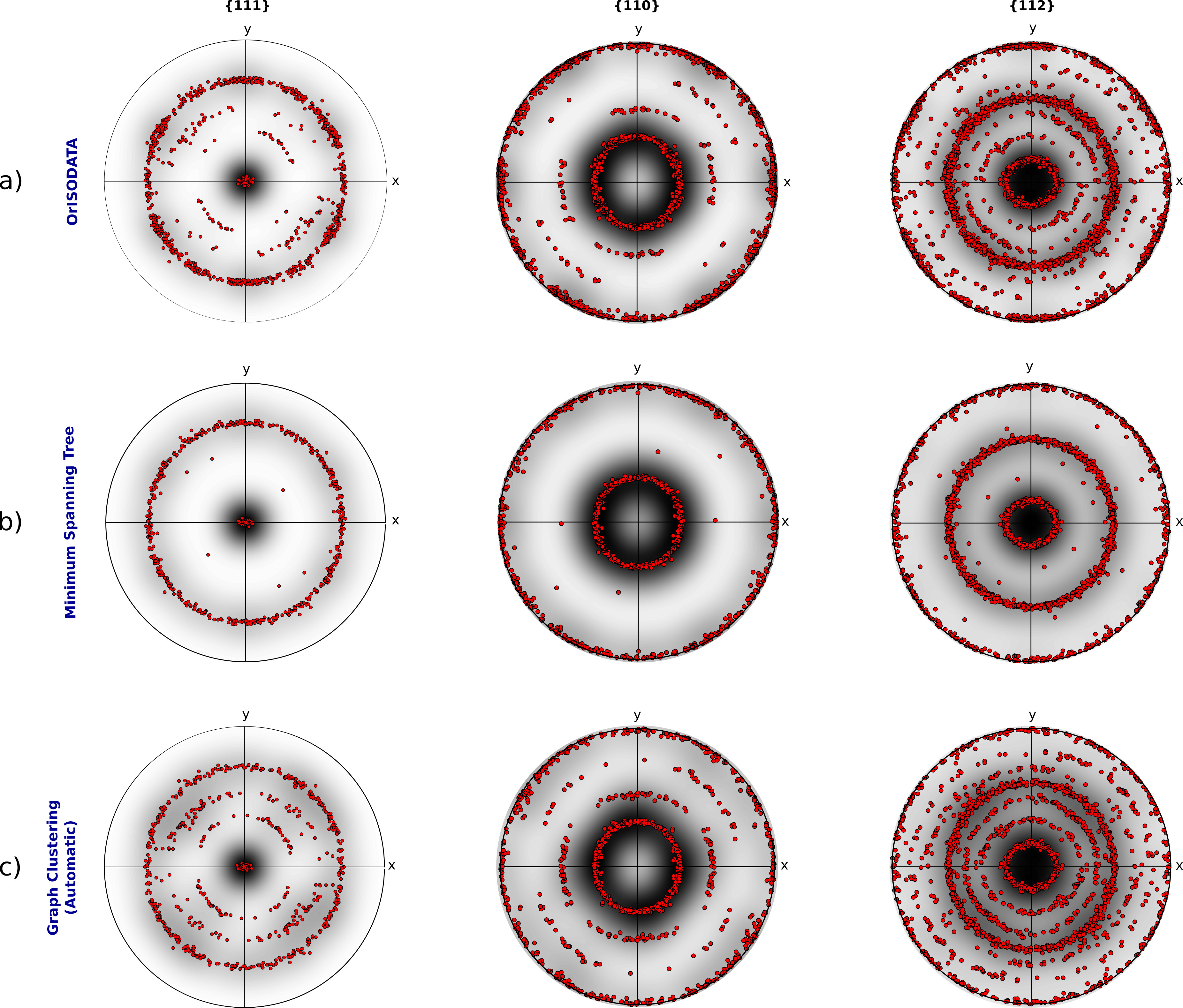}
\caption{Stereographic projections of the mean orientation of individual clusters in the deformed Au thin film sample after 10\% tensile strain, identified by a) \algo{}, b) Minimum spanning tree, and c) Graph clustering (automatic). The mean orientations are calculated using only the parent atoms, i.e. without the orphan atoms.}
\label{SuppFig:StereoProjDef}
\end{figure}

\begin{figure}[htbp!]
\centering
\includegraphics[width=0.9\textwidth]{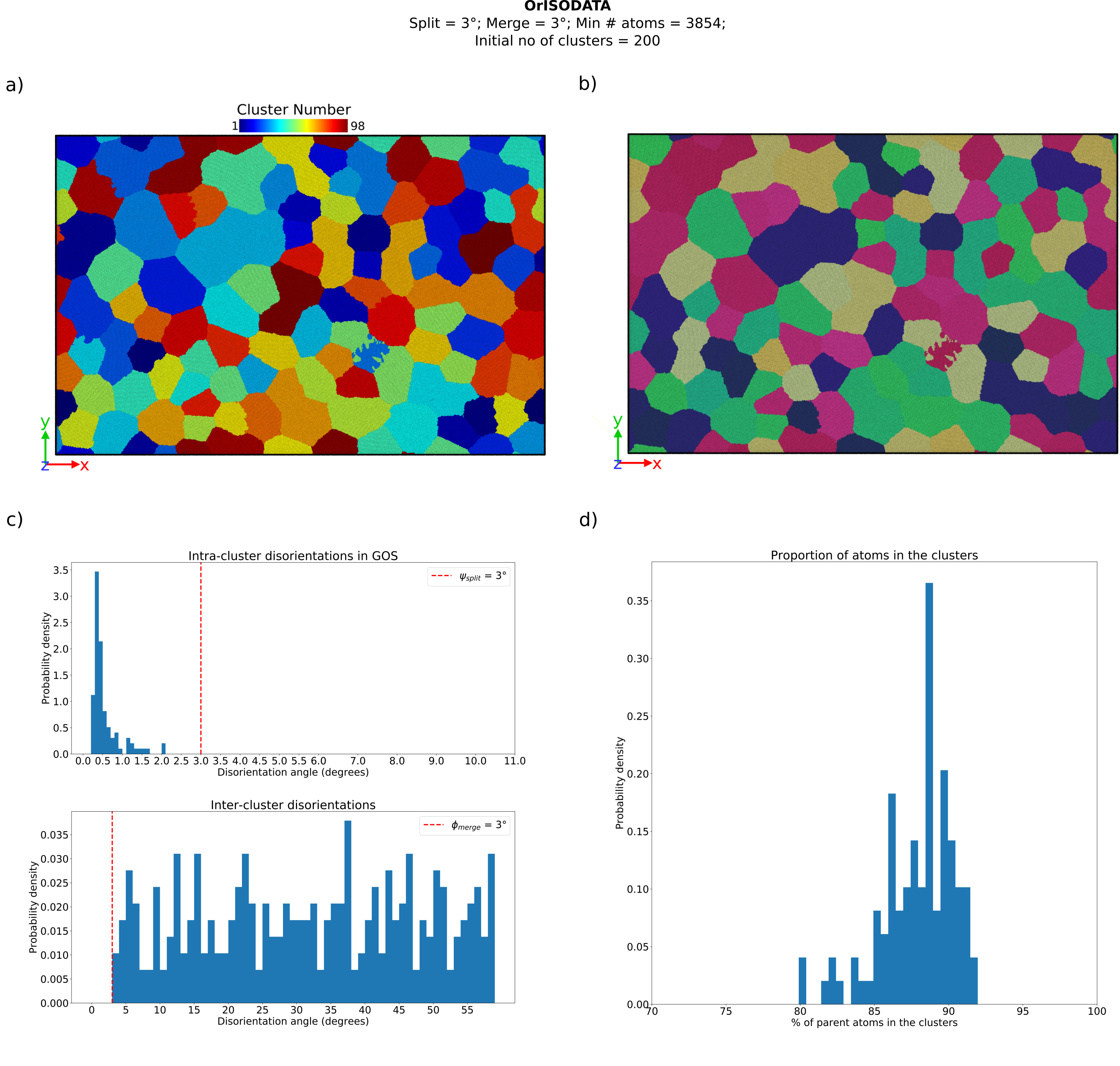}
\caption{Results from \algo{} algorithm with threshold values equivalent to the threshold value in AGC algorithm.}
\label{SuppFig:OrisodataSameThresAGC}
\end{figure}

\begin{figure}[htbp!]
\centering
\includegraphics[height=0.9\textwidth,angle=90]{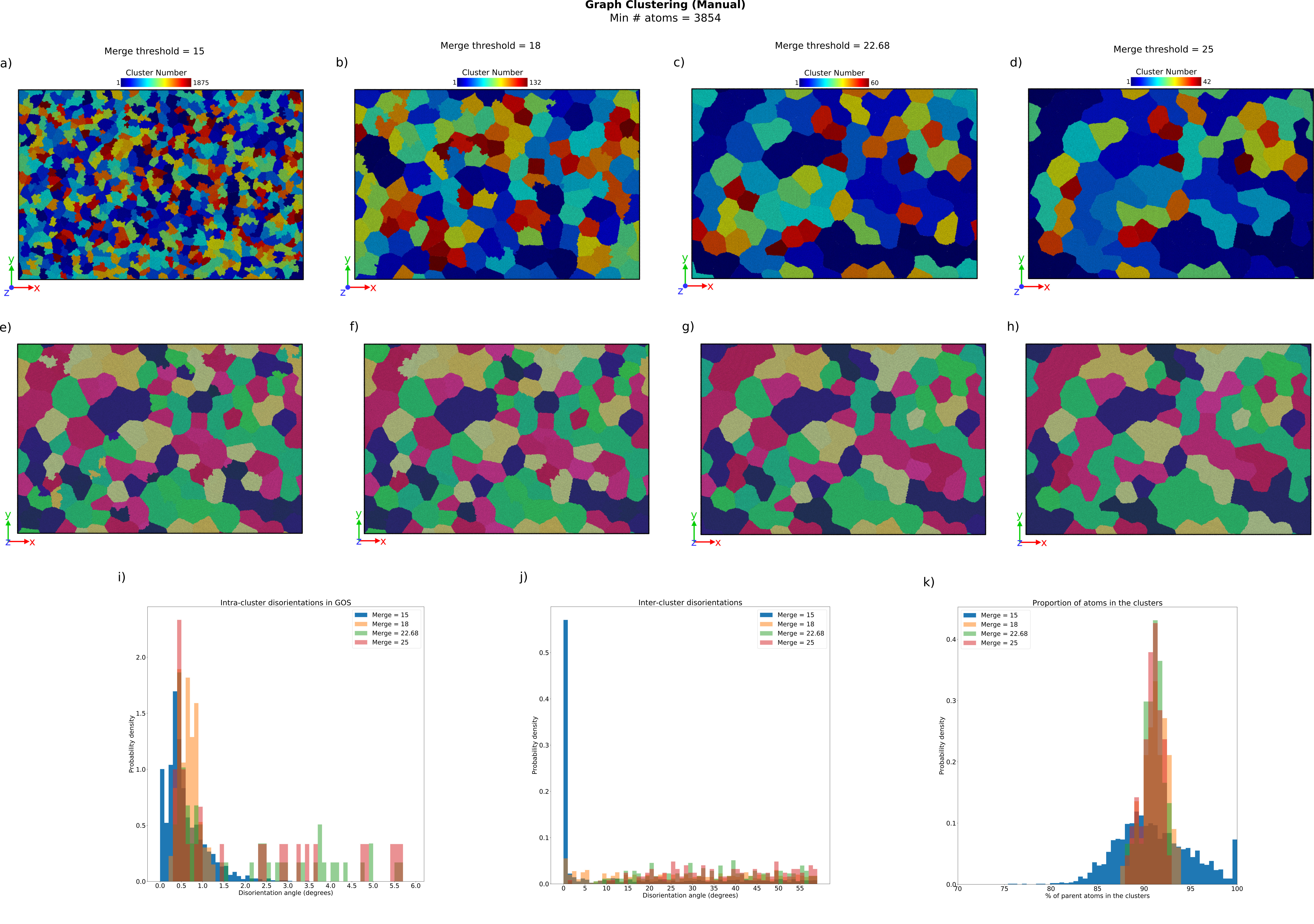}
\caption{Influence of the threshold value on the clustering results with the manual version of the graph clustering algorithm.}
\label{SuppFig:ManualGC}
\end{figure}

\section{Influence of merge threshold in OrISODATA algorithm}

\begin{figure}[htbp!]
\centering
\includegraphics[width=0.9\textwidth]{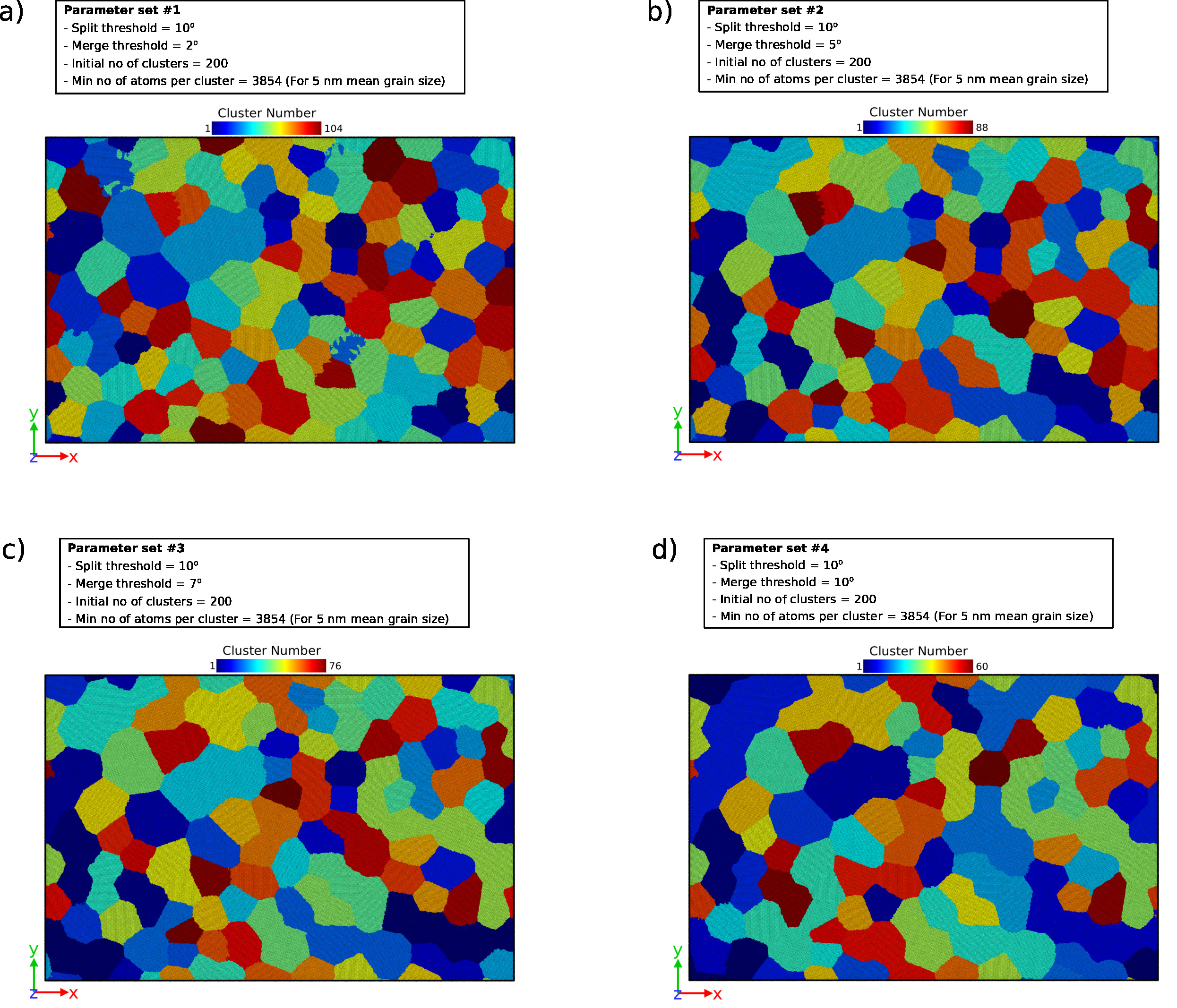}
\caption{Influence of the merge threshold on the undeformed sample case with the \algo{} clustering algorithm for merge thresholds lesser than the split threshold. Atoms are colored according to their cluster numbers.}
\label{SuppFig:MergeThresLessSplitThres}
\end{figure}

This section contains the influence of merge threshold on the undeformed
sample for the \algo{} algorithm as shown in \fref{SuppFig:MergeThresLessSplitThres}. The split threshold is kept constant and the final number of clusters decreases up on increasing merge threshold angle.

%\begin{figure}[htbp!]
%\centering
%
%\includegraphics[width=0.8\textwidth]{./Plots/supplementary/Tests_for_threshold_selection/Min_span_tree_merge_threshold_selection.eps}
%\caption{Influence of the merge threshold for the OVITO’s minimum spanning tree
%	algorithm on the final number of clusters for the strained sample case. Atoms
%	are colored according to their cluster numbers}
%\label{fig:strained_min_span_tree_merge_threshold}
%\end{figure}

%\begin{figure}[htbp!]
%\centering
%
%\includegraphics[width=0.8\textwidth]{./Plots/supplementary/Tests_for_threshold_selection/Handle_stacking_faults.eps}
%\caption{Clustering results without enabling the \emph{Handle coherent interfaces /
%		stacking faults} option for OVITO algorithms for
%	the strained sample case. Atoms are colored according to their cluster numbers}
%\label{fig:strained_no_handle_stacking_fault}
%\end{figure}

\section{Comparison of graph clustering and OrISODATA results for equal number of final clusters}

\begin{figure}[htbp!]
\centering
\includegraphics[width=0.9\textwidth]{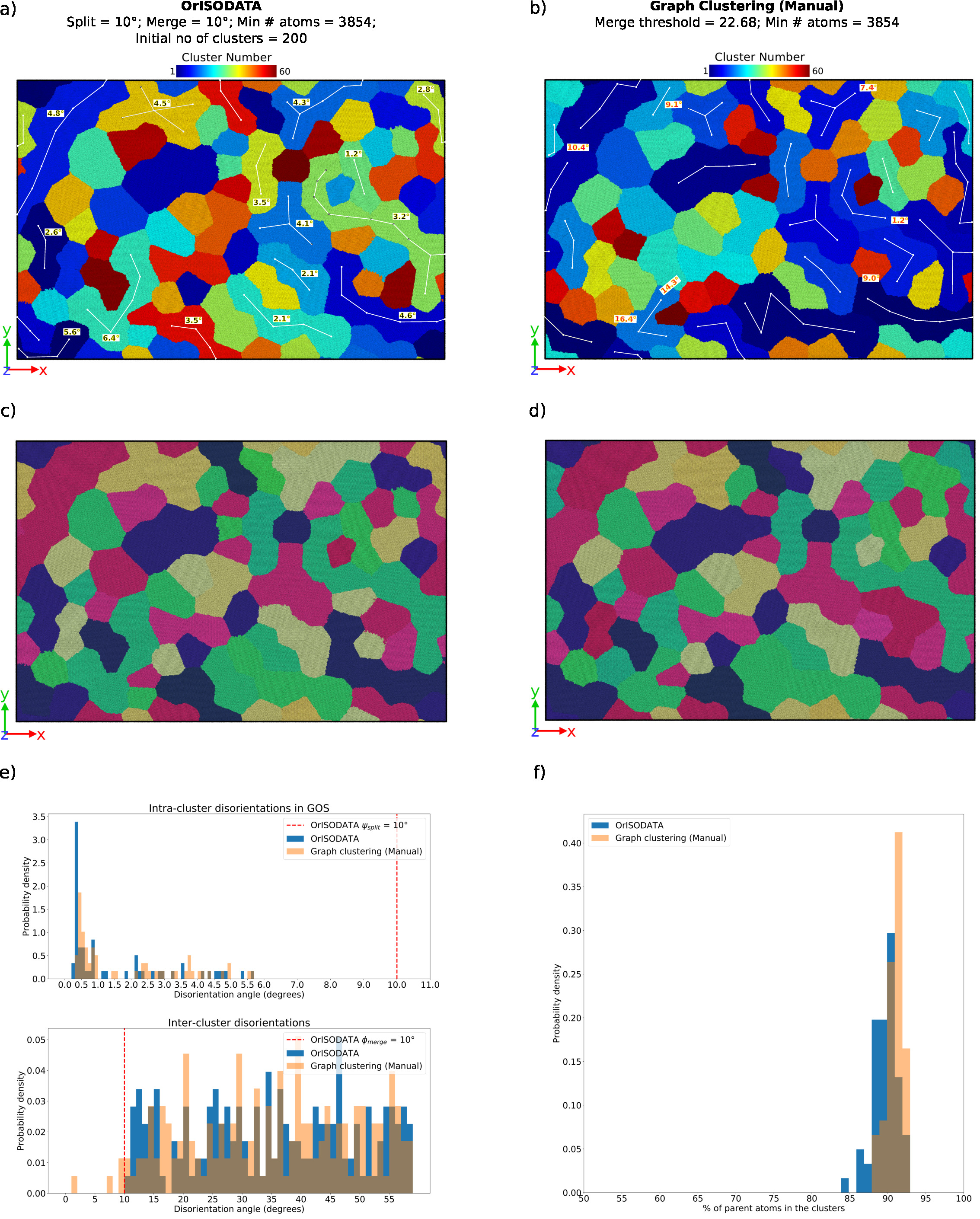}
\caption{Comparison between the graph clustering (manual) and the \algo{} results for the undeformed sample having equal number of final number of clusters. The white lines connects the regions that belongs to the same cluster number . Atoms are colored according to their cluster numbers. The angles highlighted represents the intra-cluster disorientation angles and the inter-cluster disorientation angles between certain neighbors in a) \& b) respectively. c), d) Mean orientations of clusters represented in RGB values for \algo{} and graph clustering (manual), respectively. e) Intra-cluster and inter-cluster disorientation distributions. f) Distribution of the percentage of parent atoms in the clusters comparing the graph clustering (manual) and \algo{} algorithms.}
\label{SuppFig:AGCcompareOrisodata}
\end{figure}

\Fref{SuppFig:AGCcompareOrisodata} compares the clustering results by manually tweaking the graph clustering merge threshold parameter value to get the final number of clusters equal to the number of clusters from \algo{} clustering algorithm. Most of the clusters are equivalent between the two algorithms and the intra-cluster disorientation GOS angles for the merged clusters are well below the defined threshold in the \algo{} algorithm case. Both algorithms retained a maximum proportion of parent cluster atoms (See \fref{SuppFig:AGCcompareOrisodata} f) and this explains the difference between the two algorithms in the percentage of parent atoms distribution for the undeformed sample case.

\begin{figure}[htbp!]
\centering
\includegraphics[width=0.9\textwidth]{FigsNew/SuppFig_InfluenceInitClusters.jpg}
\caption{Influence of $n^{init}_{clus}$ on the final number of clusters for the three different samples in the current work. a) Final number of clusters for varying number of initial clusters. For each test, at least three different random realizations of the initial number of clusters was used. b) Box plot showing the variation in the final number of clusters over all runs.}
\label{SuppFig:InfluInitClus}
\end{figure}

\end{document}